\documentclass[
 amsmath,
 amssymb,
 aps,
 prd,
 showpacs,
 superscriptaddress,
 nofootinbib
]{revtex4-1}

\usepackage{graphicx}
\usepackage[font=small,format=plain,labelfont=bf,up,textfont=normal,up,justification=justified,singlelinecheck=false]{caption}
\usepackage{subcaption}
\usepackage{tabularx}
\usepackage{tikz}
\usepackage{pgfplots}
\usepackage{mathrsfs}
\usepackage{mathtools}
\usepackage{soul}
\usepackage{float}
\usepackage{epstopdf}
\newcommand{\dotsize}{2pt}
\newcommand{\arcwidth}{2}
\newcommand{\gapwidth}{1}

\newcommand{\Square}[1]{+(-#1,-#1) rectangle +(#1,#1);}
\newcommand{\NodesUp}[3]{
 \fill (#2,#3) circle[radius=\dotsize];
 \draw (#2,#3) node[below]{0};
 \ifnum #1 > 0
  \foreach \x in {1, ..., #1} {
   \fill (\arcwidth*\x/#1+#2,#3) \Square{\dotsize};
   \draw (\arcwidth*\x/#1+#2,#3) node[below]{\x};
  };
 \fi 
}
\newcommand{\NodesDn}[3]{
 \fill (#2,#3) circle[radius=\dotsize];
 \draw (#2,#3) node[above]{0'};
 \ifnum #1 > 0
  \foreach \x in {1, ..., #1} {
   \fill (\arcwidth*\x/#1+#2,#3) \Square{\dotsize};
   \draw (\arcwidth*\x/#1+#2,#3) node[above]{\x'};
  };
 \fi 
}
\newcommand{\DrawLine}[6]{
 \ifnum #5 > 0
  \def\a{(\arcwidth*#5/#1+#3,#4)};
 \else
  \def\a{(-\arcwidth*#5/#2+#3,-\gapwidth+#4)};
 \fi
 \ifnum #6 > 0
  \def\b{(\arcwidth*#6/#1+#3,#4)};
 \else
  \def\b{(-\arcwidth*#6/#2+#3,-\gapwidth+#4)};
 \fi
 \draw [black,dotted] \a -- \b;
}
\newcommand{\Correlations}[8]{
 \DrawLine{#1}{#2}{#3}{#4}{#5}{#6};
 \DrawLine{#1}{#2}{#3}{#4}{#7}{#8};
 \draw [black,dotted] (#3,#4) -- (#3,-\gapwidth+#4);
}
\newcommand{\CouplingUpR}[4]{
 \NodesUp{#1}{#2}{#3};
 \foreach[count=\i] \couple in {#4} {
  \def\b{#1}
  \ifx\i\b
   \draw [black] (\arcwidth*\i/#1+#2,#3) arc (0:180:\arcwidth/#1/2*\couple);
  \else
   \draw [black] (\arcwidth*\i/#1+#2,#3) arc (0:180:\arcwidth/#1/2*\couple);
  \fi
 };
}

\newcommand{\CouplingUp}[4]{
 \NodesUp{#1}{#2}{#3};
 \ifnum #1 > 0
  \foreach[count=\i] \couple in {#4} {
   \draw (\arcwidth*\i/#1+#2,#3) arc (0:180:\arcwidth/#1/2*\couple);
  };
 \fi
}
\newcommand{\CouplingDn}[4]{
 \NodesDn{#1}{#2}{#3};
 \ifnum #1 > 0
  \foreach[count=\i] \couple in {#4} {
   \draw (\arcwidth*\i/#1+#2,#3) arc (0:-180:\arcwidth/#1/2*\couple);
  };
 \fi
}

\begin{document}

\renewcommand{\eqref}[1]{Eqn.~(\ref{#1})}

\title{Formulating Weak Lensing from the Boltzmann Equation and Application to Lens-lens Couplings}

\author{S.-C.~Su}
\affiliation{Centre for Theoretical Cosmology, Department of Applied Mathematics and Theoretical Physics,
University of Cambridge, Wilberforce Road, Cambridge CB3 0WA, United Kingdom}
\author{Eugene A.~Lim}
\affiliation{Theoretical Particle Physics and Cosmology Group,
Physics Department, Kings College London, Strand, London WC2R 2LS, United Kingdom}

\date{\today}

\begin{abstract}
The Planck mission has conclusively detected lensing of the Cosmic Microwave Background (CMB) radiation from foreground sources to an overall significance of greater than $25\sigma$ \cite{LensingPlanck}. The high precision of this measurement motivates the development of a more complete formulation of the calculation of this effect. While most effects on the CMB anisotropies are widely studied through direct solutions of the Boltzmann equation, the non-linear effect of CMB lensing is formulated through the solutions of the geodesic equation. In this paper, we present a new formalism to the calculation of the lensing effect by \emph{directly solving the Boltzmann equation}, as we did in the calculation of the CMB anisotropies at recombination. In particular, we developed a diagrammatic approach to efficiently keep track of all the interaction terms and calculate all possible non-trivial correlations to arbitrary high orders. Using this formalism, we explicitly articulate the approximations required to recover 
the usual remapping approach used in current studies of the weak lensing. In addition, we point out additional unexplored corrections that are manifest in our formalism to which experiments may be sensitive. As an example, we calculate the correction to the CMB temperature power spectrum for the \emph{lens-lens} coupling effects which are neglected in standard calculations. We find that the correction is  $\lesssim 0.1\%$ of the CMB temperature power spectrum for $\ell$ up to 3000 and thus is comparable to the cosmic variance. 

\end{abstract}

\maketitle

\section{Introduction}
CMB anisotropies generated at the last scattering surface (LSS) propagating towards the present observer are inevitably distorted by the perturbed metric of the intervening spacetime. Such distortions generate higher-order fluctuations which include \emph{lensing}, \emph{redshift} and \emph{time-delay effects}. Among these non-linear effects, weak lensing  has the most significant impact on the CMB. 

CMB lensing was first considered in 1987 \cite{FirstLensing} and, since then (for a review, see \cite{LensingReportLewis}) its contributions to the power spectra \cite{LensingSeljak,LensingMZald,LensingPSBSWHu,LensingFullLewis}, bispectra \cite{ISWLensingSpergel,LensingPSBSWHu,ISWLensingLewis} and trispectra \cite{LensingTSWHu,LensingTSCooray} of the CMB temperature and polarizations have been studied. Recently, the first detection of the CMB lensing was achieved by cross-correlating the lensing potential reconstructed by the Wilkinson Microwave Anisotropy Probe (WMAP) data and large scale structure surveys (as suggested in \cite{LensCrossIdeaHirata})  \cite{LensCrossDetectSmith,LensCrossDetectHirata}.  The detection of the power spectrum of the lensing potential was first obtained with the Atacama Cosmology Telescope (ACT) data \cite{LensingACT} and then with the South Pole Telescope (SPT) data \cite{LensingSPT}. The first full-sky map of the lensing 
potential was reconstructed by the Planck data \cite{LensingPlanck} and the significance of the detection has been boosted to larger than $25\sigma$. Planck also detected the ISW-lensing bispectrum generated by the non-Gaussian lensed CMB temperature fluctuations with significance of $2.6\sigma$ \cite{ISWLensingPlanck}. Finally, the B-mode polarization induced by lensing was detected with the SPTpol data \cite{LensingBmodeHanson} at $7.7\sigma$ significance. Driven by these sophisticated experiments, the study of the CMB lensing has entered a new era with unprecedented precision.

In particular, since its major contribution comes from lenses at low redshifts ($z\lesssim 10$), this effect is very sensitive to the late-time evolution of the universe. 
By using the non-Gaussian properties of the lensing effect on the CMB temperature and polarizations, different estimators \cite{LensReconTempHirata,LensReconHu,LensReconCooray,LensReconPolHirata} were proposed to reconstruct the lensing power spectrum from the CMB data. Measuring the lensing power spectrum with high angular resolution and high sensitivity imposes constraints \cite{MassNeutrinoSong,MassNeutrinoJulien} on the neutrino masses $m_\nu$ with $\sigma(m_\nu)$ as small as 0.035 eV \cite{MassNeutrinoJulien}, which are much tighter compared to the one using the CMB power spectra alone and have fewer biasing issues compared to the constraints from the large-scale structure. The lensing power spectrum can also break the degeneracy between the neutrino masses and the equation of state parameter $\omega$ of the dark energy \cite{MassNeutrinoSong}. The cross-correlations between the reconstructed lensing potential and other late-time observations\footnote{For example, the integrated Sachs-Wolfe  (ISW) 
effect, 
the Sunyaev-Zeldovich (SZ) effect, the Cosmic Infrared Background (CIB), galaxies and quasars.} have been analyzed through various experiments \cite{LensingPlanck,ACTLensQuasar,SPTLensGalaxy,SPTLensCIB,PlanckLensCIB}. These cross-correlations allow us to constrain the dynamics of the dark matter and the dark energy \cite{CrossLensDESeljak,CrossLensDEHu}.

In view of these applications of the CMB lensing, it is clear that an accurate computation of the lensing effect is required. In particular, it remains to be articulated the \emph{approximations} involved in the usual lensing calculation. For example, in the canonical approach \cite{LensingPSBSWHu,LensingFullLewis},
the lensed CMB temperature anisotropies $\tilde{\Theta}$ are expressed in terms of the unlensed CMB temperature anisotropies $\Theta$ in the remapping\footnote{Similar approaches apply to the CMB polarizations as well.}
\begin{equation} \label{mappingapproach}
  \tilde{\Theta}(\hat{\mathbf{n}})=\Theta(\hat{\mathbf{n}}+\boldsymbol{\alpha}),
\end{equation}
where $\hat{\mathbf{n}}$ is the direction in the sky, and the deflection angle $\boldsymbol{\alpha}$ is a perturbation defined as
\begin{equation} \label{mappingequation}
  \boldsymbol{\alpha}(\hat{\mathbf{n}}) = \nabla_{\hat{\mathbf{n}}} \psi(\hat{\mathbf{n}}) \equiv 2 \int^{r_{\text{LSS}}}_0 \text{d}r\frac{r-r_{\text{LSS}}}{r~r_\text{LSS}}\nabla_{\hat{\mathbf{n}}}\Psi_\text{W}
  (r,-r\hat{\mathbf{n}}),
\end{equation}
where $\psi(\hat{\mathbf{n}})$ is known as the \emph{lensing potential}, $\Psi_\text{W}$ is the Weyl potential, $r$ is the conformal distance between the gravitational potential and the observer, $r_\text{LSS}$ is the conformal distance between the LSS and the observer, and $\nabla_{\hat{\mathbf{n}}}$ is the covariant derivative on the sphere. The usual derivation of this equation using a perturbed geodesic employs some implicit approximations. While there are assessments about the fidelity of the corrections of the remapping approach especially in high orders \cite{LensReconPolHirata,LensCorrectionHu,LensCorrectionScott}, keeping correction terms consistently in high orders is non-trivial \cite{LensCorrectionCooray,LensCorrectionHirata} and a \emph{systematic} study in arbitrarily high orders has not been undertaken. 

In this paper, we intend to fill this gap. In our approach, instead of a remapping, we \emph{derive the lensing effect by solving the Boltzmann equation to arbitrarily high orders}. In addition to encoding the geometrical information implicit in the use of the geodesic equations, our approach using the Boltzmann equation also manifestly contains additional interactions, such as redshift effects and Compton scattering. We will demonstrate how these effects couple with lensing in high orders. We will explicitly re-derive \eqref{mappingequation} and identify the implicit approximations used in the remapping approach. One of the primary benefits of our approach is that the meaning of each physical term is clear and unambiguous. Although the focus of this paper is to establish the formalism and we leave the quantitative assessments of the approximations in the future, we will show how our approach facilitates the identification of the dominant effects in high  orders. As an application, we calculate the 
corrections to the temperature power spectrum from lens-lens couplings in Section \ref{ValidApprox}.

Our paper is organized as follows. In Section \ref{SecFormalism}, we perturb the metric and express the $N$th-order intensity matrix of the CMB photons as a time integration of the $N$th-order source function. The expression is valid as long as the perturbation theory holds. We then demonstrate the formalism in 2nd order and identify the approximation needed in order to recover the remapping approach in 2nd order. Finally, we generalize the formalism to arbitrarily high orders and again recover the remapping approach. In Section \ref{ValidApprox}, we present a systematic and diagrammatic approach to represent all possible coupling terms in high orders. Focusing on the corrections from lens-lens couplings, we illustrate how these diagrams can facilitate the calculation of the lensing effects on the CMB power spectra. We discuss some limitations and possible extensions of the current work in Section \ref{SecDiscussion} and conclude in Section \ref{SecConclusions}. 

\section{Formalism}\label{SecFormalism}
In this section, we construct the coordinate system for the manifold and the tetrad basis for the tangent planes used in this paper. The construction allows us to expand the perturbations consistently to any orders. We then define the intensity matrix which embeds the intensity and the polarizations of the CMB photons. After that, we derive the weak lensing effect from the 2nd-order Boltzmann equation and generalize the derivation to arbitrarily high orders to establish the complete formalism. Finally, we recover the remapping approach of the CMB lensing in literature from our formalism by identifying the approximations required. 

\subsection{Coordinate System and Tetrad Basis}
The homogeneity and isotropy of the universe allow us to map the physical manifold onto a background manifold equipped with a Friedmann metric $g_{\mu\nu}$. By assuming a flat universe, we choose the coordinate system $\{x^A\}=\{\eta, x^I\}$ for $A=0,1,2,3$ and $I=1,2,3$ such that
\begin{eqnarray}\label{Metric}
 ds^2=g_{\mu\nu}dx^{\mu}dx^{\nu}
     =a^2(\eta)\left\{-(1+2\Phi)d\eta^2+2\mathscr{B}_I dx^I d\eta+\left[(1-2\Psi)\delta_{IJ}+2\mathscr{H}_{IJ}\right]dx^I dx^J\right\},
\end{eqnarray}
where $a(\eta)$ is the scale factor, and $\Phi$, $\Psi$, $\mathscr{B}_I$ and $\mathscr{X}_{IJ}$ are perturbations of the metric as functions of $x^A$. The coordinates $x^A$ label the background manifold -- the Greek indices $(\mu,\nu,\rho,\dots)$ are abstract indices and do not run\footnote{A more familiar notation would be to write the metric in \eqref{Metric} as $g_{AB}dx^Adx^B$. That is, the indices $A$ and $B$ do double duty as coordinate labels and abstract indices.}. 
All the metric perturbations in \eqref{Metric} can be considered as fields living on the background manifold. We further decompose $\mathscr{B}_I$ and $\mathscr{X}_{IJ}$ by using the scalar-vector-tensor decomposition such that
\begin{eqnarray}
 \mathscr{B}_I&=&\partial_I\mathcal{B}+\mathcal{B}_I,\\
 \mathscr{H}_{IJ}&=&\mathcal{H}_{IJ}+\partial_{(I} \mathcal{E}_{J)}+\partial_I\partial_J \mathcal{E},
\end{eqnarray}
where $2\partial_{(I} \mathcal{E}_{J)}\equiv\partial_{I} \mathcal{E}_{J}+\partial_{J} \mathcal{E}_{I}$ and $\partial^I\mathcal{B}_I=\partial^I\mathcal{E}_I=\partial^I\mathcal{H}_{IJ}=0$ and ${\mathcal{H}^I}_I=0$ with $\partial_I\equiv\partial/\partial x^I$. Throughout this paper, we use the Newtonian gauge, i.e.~
\begin{eqnarray}
 \mathcal{E}=\mathcal{B}=\mathcal{E}_I=0.
\end{eqnarray}
With the assumption of a flat Friedmann metric in the background order, the equations above hold in general and can be expanded into perturbations of different orders in the following way
\begin{equation}
 \mathcal{W}=\mathcal{W}^{[\text{I}]}+\frac{\mathcal{W}^{[\text{II}]}}{2!}+\frac{\mathcal{W}^{[\text{III}]}}{3!}\hdots,
\end{equation}
where the Roman numbers inside the square brackets of the superscripts denote the orders of perturbations.

At each point on the physical manifold, we construct the tangent basis to decompose the photon momenta. It is convenient to introduce the tetrad fields $\hat{\mathbf{e}}_a$ which satisfy the orthonormality conditions
\begin{eqnarray}
 \hat{\mathbf{e}}_a\cdot\hat{\mathbf{e}}_b\equiv g_{\mu\nu} {e_a}^\mu {e_b}^\nu = \eta_{ab},\nonumber\\
 \hat{\mathbf{e}}^a\cdot\hat{\mathbf{e}}^b\equiv g^{\mu\nu} {e^a}_\mu {e^b}_\nu = \eta^{ab},
\end{eqnarray}
where $\eta_{ab}$ is the Minkowski metric, and $a$ and $b$ run from 0 to 3 labeling the four tetrad vectors and forms. Since photons have null geodesics ($p^\mu p_\mu=0$), their momenta can be written as
\begin{equation} \label{pequation}
 \mathbf{p}=p^0(\hat{\mathbf{e}}_0+\hat{\mathbf{n}}),
\end{equation}
where $p^0$ is the photon energy measured by an observer with the velocity $\hat{\mathbf{e}}_0$ and $\hat{\mathbf{n}}$ is a spacelike vector on the hypersurface ``screen'' denoting the direction of the photon as seen by the observer, such that  $n_\mu {e_0}^\mu=0$ and $n^\mu n_\mu=1$.  Furthermore, it is useful to define the screen projector
\begin{equation}
 S_{\mu\nu}(\hat{\mathbf{e}}^0,\mathbf{p})\equiv g_{\mu\nu} + {e^0}_\mu {e^0}_\nu - n_\mu n_\nu.
\end{equation}
By foliating spacetime into hypersurfaces threaded through by the orbit defined by the observer's velocity $\hat{\mathbf{e}}_0$, we can describe radiation by a screen-projected rank-2 tensor -- the intensity matrix which lives on these hypersurfaces. We will discuss the intensity matrix in detail in Section \ref{SubSecIntensity}.

We note that the tetrad fields $\hat{\mathbf{e}}_a$ lie on the physical manifold. Thus, we have to pull them back onto the background manifold in order to study the Boltzmann equation on the background manifold. This is achieved \cite{NthOrderPerturbations} through the pullback $\phi^*$ with a gauge field $\xi$ defined by 
\begin{equation}\label{pullback_e}
 {_\xi}\hat{\mathbf{e}}_a\equiv \lim_{\lambda\rightarrow 1}\phi^*_{\lambda,\xi}(\hat{\mathbf{e}}_a)=\lim_{\lambda\rightarrow 1}\sum_{N=0}^\infty\frac{\lambda^N}{N!}\mathcal{L}_\xi^N\hat{\mathbf{e}}_a,
\end{equation}
where $\lambda$ denotes the foliations $\mathcal{M}_\lambda$ of an embedding (4+1)-dimensional manifold $\mathcal{N} = \mathcal{M} \times [0, 1]$\footnote{We define the background manifold as $\mathcal{M}_0$ and the physical manifold as $\mathcal{M}_1$ with $\lambda\in[0,1]$.} and $\mathcal{L}$ is the Lie derivative. The pulled-back tetrads ${_\xi}\hat{\mathbf{e}}_a$ lie on the background manifold. Using the following natural background basis for vectors and forms on the background manifold

\begin{equation}
 \mathbf{\bar{e}}_A\equiv\frac{1}{a(\eta)}\frac{\partial}{\partial x^A}~,~~~
 \mathbf{\bar{e}}^A\equiv a(\eta)\text{d}x^A,
\end{equation}
we can express the tetrads as (with the alignment  ${_\xi}\mathbf{\bar{e}}_a = \mathbf{\bar{e}}_A$)
\begin{eqnarray}
 {_\xi}\hat{\mathbf{e}}_a = {_\xi}{X^b}_a \mathbf{\bar{e}}_b~,~~~
 {_\xi}\hat{\mathbf{e}}^a = {_\xi}{Y^a}_b \mathbf{\bar{e}}^b,
\end{eqnarray}
where the coefficients can be expanded into perturbations as
\begin{eqnarray}
 {_\xi}X_{ab}=\sum_{N=0}^\infty\frac{\lambda^N}{N!}{_\xi}X_{ab}^{[N]}~,~
 {_\xi}Y_{ab}=\sum_{N=0}^\infty\frac{\lambda^N}{N!}{_\xi}Y_{ab}^{[N]}.
\end{eqnarray}
We can calculate the coefficients order by order using 
\begin{equation}
 \eta_{ab}=\phi^*_{\lambda,\xi}(\eta_{ab})=\phi^*_{\lambda,\xi}({e_a}^A) \phi^*_{\lambda,\xi}({e_b}^B) \phi^*_{\lambda,\xi}(g_{AB}).
\end{equation}
There remain residual freedoms in determining the anti-symmetric part of the coefficients ${_\xi}X_{ab}$ and ${_\xi}Y_{ab}$ due to boost and rotation. In particular, we align ${_\xi}\hat{\mathbf{e}}^0$ with $\text{d}\eta$, i.e.~${_\xi} Y_{i0}=0$. Physically, it means that the velocity of the chosen observer is orthonormal to the hypersurfaces of constant time. This fixes the boost freedom such that ${_\xi}Y_{[i0]}=-{_\xi}Y_{[0i]}=-{_\xi}Y_{(i0)}$ with $(\hdots)$ and $[\hdots]$ denoting the symmetric and anti-symmetric parts. Finally, the rotation freedom can be fixed by setting ${_\xi}Y_{[ij]}=0$.

\subsection{Intensity Matrix}\label{SubSecIntensity}
To encode all the information about the intensity and the polarization of the CMB photons, we define the screen-projected intensity matrix $\mathscr{P}_{\mu\nu}(x^A,p^a)$ which satisfies\footnote{For details about how to construct the intensity matrix, see \cite{IntensityMatrix}.}.
\begin{equation}
  e_0^\mu \mathscr{P}_{\mu\nu}(x^A,p^a) 
 = n^\mu \mathscr{P}_{\mu\nu}(x^A,p^a) = p^\mu \mathscr{P}_{\mu\nu}(x^A,p^a) = 0.
\end{equation}

The intensity matrix of the CMB radiation can be decomposed into
\begin{equation}\label{intensitymatrixdecompose}
 \mathscr{P}_{\mu\nu} = \frac{1}{2} \mathcal{I} S_{\mu\nu} + \mathcal{P}_{\mu\nu},
\end{equation}
where $\mathcal{I}$ denotes the photon intensity, and the symmetric and trace-free polarization tensor $\mathcal{P}_{\mu\nu}$ encodes the linear polarizations of the CMB photons. Here, we ignore the circular polarization because it is not induced by Compton scattering and thus is usually neglected in the CMB studies. The polarization tensor can be further decomposed into two scalar fields -- E-mode  $\mathcal{P}_E$ and B-mode $\mathcal{P}_B$, i.e. \cite{LensingReportLewis}
\begin{equation}
 \mathcal{P}_{\mu\nu} = \tilde{\nabla}_{\langle\mu} \tilde{\nabla}_{\nu\rangle} \mathcal{P}_E + {\epsilon^\gamma}_{(\mu}\tilde{\nabla}_{\gamma)}\tilde{\nabla}_\nu \mathcal{P}_B,
\end{equation}
where $\tilde{\nabla}_\mu \equiv {S_\mu}^\nu \nabla_\nu$ denotes the screen-projected covariant derivative, $( \hdots )$ denotes the symmetric part, $\langle\hdots\rangle$ denotes the symmetric trace-free part, and $\epsilon^{\mu\nu}\equiv-i(e^\mu_{-}e^\nu_{+} - e^\nu_{-}e^\mu_{+})/2$ with $\mathbf{e}_\pm\equiv\hat{\mathbf{e}}_x\pm i\hat{\mathbf{e}}_y$\footnote{$\hat{\mathbf{e}}_x$ and $\hat{\mathbf{e}}_y$, which are orthogonal to each other, lie on the plane perpendicular to the direction $\hat{\mathbf{n}}$ of the light path and the observer's velocity $\hat{\mathbf{e}}_0$.}.

The normalized energy-integrated photon intensity and polarization tensor are defined as
\begin{equation}\label{EnergyIntI}
 \hat{I}(x^A,\hat{\mathbf{n}})\equiv\frac{1}{\bar{I}(\eta)}\int \mathcal{I}(x^A,p^0,\hat{\mathbf{n}}) (p^0)^3 \text{d} p^0
\end{equation}
and
\begin{equation}\label{EnergyIntP}
 \hat{P}_{\mu\nu}(x^A,\hat{\mathbf{n}})\equiv\frac{1}{\bar{I}(\eta)}\int \mathcal{P}_{\mu\nu}(x^A,p^0,\hat{\mathbf{n}}) (p^0)^3 \text{d} p^0,
\end{equation}
where $\bar{I}(\eta)\equiv\int \bar{\mathcal{I}}(\eta,p^0) (p^0)^3 \text{d} p^0$ with $\bar{\mathcal{I}}(\eta,p^0)$ as the homogeneous and isotropic black-body spectrum in the background order.

\subsection{Boltzmann Equation}
We are ready to formulate the Boltzmann equation. From now on, we stay with the tetrad basis (denoted by indices $a,b,c,\hdots$) and elaborate all relevant equations onto the background manifold. The Boltzmann equation consists of two operators, the Liouville operator $\mathfrak{L}$ describing the free-streaming of photons and the collision operator $\mathfrak{C}$ describing Compton scattering. That is,
\begin{equation}\label{BoltzmannEqn}
 \mathfrak{L}[\mathscr{P}_{ab}(x^A,p^0,\hat{\mathbf{n}})]=\mathfrak{C}_{ab}(x^A,p^0,\hat{\mathbf{n}}),
\end{equation}
where we have switched to the tetrad basis via
\begin{equation}
  \mathscr{P}_{ab} = \mathscr{P}_{\mu\nu}{e_a}^{\mu} {e_b}^{\nu}~,~~~\mathfrak{C}_{ab} = \mathfrak{C}_{\mu\nu} {e_a}^{\mu} {e_b}^{\nu},~~~S_{ab} = S_{\mu\nu} {e_a}^{\mu} {e_b}^{\nu}.
\end{equation}
We remark that the Boltzmann equation as a whole is gauge-invariant.

We begin with the Liouville operator, which is defined as 
\begin{equation}\label{Liouville}
 \mathfrak{L}\equiv{S_a}^c {S_b}^d ~\frac{\text{d}}{\text{d}s}={S_a}^c {S_b}^d \left[ p^A\nabla_A + \frac{\text{d}p^e}{\text{d}s}\partial_{p^e}\right],
\end{equation}
where $\nabla_A$ is the covariant derivative with respect to $x^A$, $\partial_{p^e}\equiv\partial/\partial p^e$ and $p^A\equiv\text{d}x^A/\text{d}s$. In this paper, we choose the conformal time $\eta$ as the affine parameter instead of the proper distance $s$. We can do so by multiplying \eqref{BoltzmannEqn} with the Jacobian $\text{d}s/\text{d}\eta$. Linearity of the Liouville operator preserves the decomposition of $\mathscr{P}_{ab}$ in \eqref{intensitymatrixdecompose}, i.e.~
\begin{equation}
 \mathfrak{L}[\mathscr{P}_{ab}] = \frac{1}{2} \mathfrak{L}[\mathcal{I}] S_{ab} + \mathfrak{L}[\mathcal{P}_{ab}].
\end{equation}
That means we can write down the evolution equations of the intensity and polarizations of photons separately\footnote{However, the equations are coupled through Compton scattering.}.

In practice, we perturb the Boltzmann equation and the Einstein field equations, and solve these differential equations order by order. As usual, quantities in lower orders are treated as sources (or inhomogeneous part) of the higher-order differential equations. In other words, we start from the lowest (background) order and iteratively solve the equations order by order until we get the precision required.

The photon trajectory is given by the geodesic equation
\begin{equation}
 \frac{\text{d}p^a}{\text{d}s}+{{\omega_b}^a}_c~p^bp^c = 0,
\end{equation}
where the Ricci rotation coefficients are given by
\begin{equation}
{{\omega_b}^a}_c\equiv{e^a}_\mu{e_b}^\nu\nabla_\nu{e_c}^\mu=\Gamma^I_{JK} {e_b}^K {e_c}^J {e^a}_I + {e_b}^I {e^a}_J \partial_I {e_c}^J,
\end{equation}
which are determined completely by the metric $g_{AB}$ and the tetrads ${e_a}^A$. Meanwhile, the momentum $p^A$ can be expressed in the tetrad basis
\begin{equation}
 p^A=p^a{e_a}^A.
\end{equation}
We can expand the quantities into perturbations as
\begin{eqnarray}
 \left(\frac{\text{d}p^a}{\text{d}s}\right)^{[N]} &=& - {{{\omega_b}^a}_c}^{[N]}~p^bp^c,~~~~~~~~~
 (p^A)^{[N]} = p^a ({e_a}^A)^{[N]},\\
 \mathscr{P}_{ab} &=& \bar{\mathscr{P}}_{ab}+\mathscr{P}_{ab}^{[\text{I}]}+\frac{1}{2!}\mathscr{P}_{ab}^{[\text{II}]}+\frac{1}{3!}\mathscr{P}_{ab}^{[\text{III}]}+\hdots,
\end{eqnarray}
and keep the perturbation order consistently for the Liouville term in \eqref{Liouville}. We list out the leading order terms in the following\footnote{We ignore the vector and tensor perturbations in 1st order.}
\begin{eqnarray}
 \left(\frac{\text{d} x^I}{\text{d}\eta}\right)^{[\text{0}]}&=&n^i,~~~~~~~~
 \left(\frac{\text{d} x^I}{\text{d}\eta}\right)^{[\text{I}]}=n^i(\Psi^{[\text{I}]}+\Phi^{[\text{I}]}),\\ \label{dxdetaI}
 \left(\frac{\text{d} n^i}{\text{d}\eta}\right)^{[\text{0}]}&=&0,~~~~~~~~~~
 \left(\frac{\text{d} n^i}{\text{d}\eta}\right)^{[\text{I}]}=-S^{ij}\partial_J(\Psi^{[\text{I}]}+\Phi^{[\text{I}]}),\\ \label{dndetaI}
 \left(\frac{\text{d} p^0}{\text{d}\eta}\right)^{[\text{0}]}&=&-\mathcal{H}p^0,~~~~ 
 \left(\frac{\text{d} p^0}{\text{d}\eta}\right)^{[\text{I}]}= p^0(\partial_\eta \Psi^{[\text{I}]}-n^i\partial_I\Phi^{[\text{I}]}),\\ \label{dpdetaI}
 \left(\frac{\text{d} p^0}{\text{d}\eta}\right)^{[\text{II}]}&=& p^0\left[\partial_\eta\Psi^{[\text{II}]}\right.-\left.n^i \partial_I\Phi^{[\text{II}]}+ 
 (\partial_I \mathcal{B}_J^{[\text{II}]}-\partial_\eta \mathcal{H}_{IJ}^{[\text{II}]})n^i n^j+ 
 2(\Phi^{[\text{I}]}-\Psi^{[\text{I}]})n^i\partial_I\Phi^{[\text{I}]}+4\Psi^{[\text{I}]}\partial_\eta\Psi^{[\text{I}]}\right], \label{dpdetaII}
\end{eqnarray}
where $\mathcal{H}(\eta)$ is the conformal Hubble parameter. Here, we decompose the photon momentum $\mathbf{p}$ into $p^0$ and $n^i$ using the tetrad basis as it lives on the tangent plane. Similarly, the screen projector is represented in tetrad basis, i.e.~$S^{ij}$ instead of $S^{IJ}$. In contrast, we express everything else in the coordinate system $\{x^A\}$ of the background manifold, such as metric perturbations ($\mathcal{B}_J$, $H_{IJ}$, ...) as well as the spatial derivatives ($\partial / \partial x^I$). For example, $S^{ij} \partial_J$ and $n^i\partial_I$ in the above equations stand for ${{_\xi}\bar{e}_j}^J S^{ij} \partial_J$ and ${{_\xi}\bar{e}_i}^I n^i \partial_I$ with the alignment ${_\xi}\mathbf{\bar{e}}_a=\mathbf{\bar{e}}_A$.\footnote{We use this expression because of the following reason: The photon momentum $\mathbf{p}$ at a point of the physical manifold is measured as the quantities $p^0$ and $n^i$ under a chosen basis (tetrad basis in our case) of the tangent plane. These quantities are 
unperturbed. On the other hand, the metric perturbations are expanded on the background manifold. To be consistent, we perform all the calculations on the background manifold. That is, we pull back the tetrad fields $\hat{\mathbf{e}}_a$ onto the background manifold as shown in \eqref{pullback_e}. Due to the existence of perturbations, we do not expect the pulled-back tetrads $_\xi\hat{\mathbf{e}}_a$ to be perfectly aligned with the natural background basis $\bar{\mathbf{e}}_A$. Thus, ${{_\xi}\hat{\mathbf{e}}_a}^A$ are the coefficients of the pulled-back tetrads under the natural background basis $\bar{\mathbf{e}}^A$ and are perturbative.} We omit $\mathbf{\bar{e}}$ for simplicity throughout this paper.

The same expansion can be done for the collision terms $\mathfrak{C}_{ab}$\footnote{The complete derivation of the collision terms up to 2nd order can be found in \cite{2ndBoltzmannPitrou}.}; but we will focus on the damping term $-\bar{\tau}'\mathscr{P}_{ab}$ in Compton scattering with the residual terms collected in $\mathfrak{D}_{ab}$, i.e.~
\begin{equation} \label{CollisionTerm} 
 \mathfrak{C}_{ab}(x^A,p^0,\hat{\mathbf{n}}) = - \bar{\tau}'(\eta)\mathscr{P}_{ab}(x^A,p^0,\hat{\mathbf{n}}) + \mathfrak{D}_{ab}(x^A,p^0,\hat{\mathbf{n}}),
\end{equation}
where $\bar{\tau}'(\eta)\equiv \sigma_T a(\eta)\bar{n}_e(\eta)$ is the differential optical depth in the background order with the Thomson cross section $\sigma_T$ and the mean number density $\bar{n}_e$ of free electrons. The term $-\bar{\tau}'\mathscr{P}_{ab}$ is responsible for the damping effect on the CMB anisotropies. The $N$th-order residual term $\mathfrak{D}_{ab}^{[N]}$ contains only contributions from the low multipoles (at most $\ell=2$) of the $N$th-order intensity matrix and cross terms of lower-order perturbations. For example, the low multipoles contribute to intrinsic intensity and Doppler effect to $N$th order. These low multipoles can be calculated through the Boltzmann equation and the Einstein field equations in $N$th order with a truncated value of $\ell$. Then, they are fed back into the Boltzmann equation, which we will then solve using the line of sight approach. In principle, we also need to know the perturbed differential optical depth $\delta\tau'(x^A)$ but this is out of the 
scope of this paper\footnote{We consider only the 1st-order $\mathfrak{D}_{ab}^\text{[I]}$. See Approximation \ref{ApproCol} in Section \ref{HighOrderLensing} for details.}. Nevertheless, we emphasize that there is no  dependence of $\mathscr{P}_{ab}^{[N]}$ inside the residual term $\mathfrak{D}_{ab}^{[N]}$ so it can be treated as a source term in the evolution equations of $\mathscr{P}_{ab}^{[N]}$. 

Putting together both the Liouville term \eqref{Liouville} and the collision term \eqref{CollisionTerm}, we can formulate the solution of the $N$th-order intensity matrix $\mathscr{P}_{ab}^{[N]}$ by using the line of sight approach as follows. The Boltzmann equation in $N$th order can be written as
\begin{eqnarray}\label{NthBoltzmann}
 &~&{S_a}^c {S_b}^d\left\{\frac{\partial\mathscr{P}_{cd}^{[N]}}{\partial\eta}
 +\sum_{L=0}^{N}\dbinom{N}{L}\left[
  \left(\frac{\text{d}x^I}{\text{d}\eta}\right)^{[L]}\frac{\partial\mathscr{P}_{cd}^{[N-L]}}{\partial x^I}
 +\left(\frac{\text{d}p^0}{\text{d}\eta}\right)^{[L]}\frac{\partial\mathscr{P}_{cd}^{[N-L]}}{\partial p^0}
 +\left(\frac{\text{d}n^i}{\text{d}\eta}\right)^{[L]}\frac{\partial\mathscr{P}_{cd}^{[N-L]}}{\partial n^i}
 \right]\right\}\nonumber\\
 &=&- \bar{\tau}'\mathscr{P}_{ab}^{[N]} + \mathfrak{D}_{ab}^{[N]},
\end{eqnarray}
where $\binom{N}{L}$ is the Binomial coefficient.
We rearrange \eqref{NthBoltzmann} by putting all the terms with $\mathscr{P}_{ab}^{[N]}$ on the L.H.S., such that
\begin{eqnarray}\label{NthPabPDE}
\frac{\partial\mathscr{P}_{ab}^{[N]}}{\partial\eta}
 +n^i\frac{\partial\mathscr{P}_{ab}^{[N]}}{\partial x^I}
 -\mathcal{H}p^0\frac{\partial\mathscr{P}_{ab}^{[N]}}{\partial p^0}
 + \bar{\tau}'\mathscr{P}_{ab}^{[N]}
 ={S_a}^c {S_b}^d \mathcal{Q}_{cd}^{[N]},
\end{eqnarray}
where we have collected all the source terms in $\mathcal{Q}_{ab}^{[N]}$ which is 
\begin{eqnarray}\label{sourcetermQ}
 \mathcal{Q}_{ab}^{[N]}
 \equiv
 \mathfrak{D}_{ab}^{[N]}
 -\sum_{L=1}^{N}\dbinom{N}{L}\left[
 \left(\frac{\text{d}x^I}{\text{d}\eta}\right)^{[L]}\frac{\partial\mathscr{P}_{ab}^{[N-L]}}{\partial x^I}
 +\left(\frac{\text{d}p^0}{\text{d}\eta}\right)^{[L]}\frac{\partial\mathscr{P}_{ab}^{[N-L]}}{\partial p^0}
 +\left(\frac{\text{d}n^i}{\text{d}\eta}\right)^{[L]}\frac{\partial\mathscr{P}_{ab}^{[N-L]}}{\partial n^i}
 \right].~~~~~
\end{eqnarray}
Performing the Fourier transform such that any function of $\mathbf{x}$ transforms as
\begin{eqnarray}
 f(\mathbf{x})=\int \frac{\text{d}\mathbf{k}}{(2\pi)^{\frac{3}{2}}}e^{i \mathbf{k}\cdot\mathbf{x}}f(\mathbf{k}),
\end{eqnarray}
we obtain
\begin{eqnarray} \label{EOMfirstorder}
 \frac{\partial\hat{\mathscr{P}}_{ab}^{[N]}}{\partial\eta}+i\mathbf{k}\cdot\hat{\mathbf{n}} \hat{\mathscr{P}}_{ab}^{[N]} + \bar{\tau}' \hat{\mathscr{P}}_{ab}^{[N]} = {S_a}^c {S_b}^d \hat{\mathcal{Q}}_{cd}^{[N]},
\end{eqnarray}
where $\hat{\mathscr{P}}_{ab}$ and $\hat{\mathcal{Q}}_{ab}$ are functions of $\eta$, $\mathbf{k}$ and $\hat{\mathbf{n}}$, and are normalized over energy (as in \eqref{EnergyIntI} and \eqref{EnergyIntP}) 
\begin{equation}
\hat{\mathscr{P}}_{ab}\equiv \frac{1}{\bar{I}}\int \mathscr{P}_{ab} (p^0)^3 \text{d}p^0~,~\hat{\mathcal{Q}}_{ab}\equiv \frac{1}{\bar{I}}\int \mathcal{Q}_{ab} (p^0)^3 \text{d}p^0 .
\end{equation}

Integrating \eqref{EOMfirstorder} over the line of sight, the solution to the $N$th-order intensity matrix at present ($\eta=\eta_0$) is
\begin{eqnarray}\label{PabNow}
 \hat{\mathscr{P}}_{ab}^{[N]}(\eta_0,\mathbf{k},\hat{\mathbf{n}})=\int^{\eta_0}_0 \text{d}\eta e^{-i\mathbf{k}\cdot\hat{\mathbf{n}} r -\bar{\tau}} {S_a}^c {S_b}^d \hat{\mathcal{Q}}_{cd}^{[N]}(\eta,\mathbf{k},\hat{\mathbf{n}}),
\end{eqnarray}
where the background-order optical depth $\bar{\tau}(\eta)\equiv \int^{\eta_0}_{\eta}\bar{\tau}'(\tilde{\eta})d\tilde{\eta}$ and the conformal distance $r\equiv\eta_0-\eta$. We remark that all the quantities in $\hat{\mathcal{Q}}_{ab}^{[N]}$ can be determined either by solving the Einstein field equations and the Boltzmann equation with a low truncated $\ell$ for up to $N$th order or by performing the line of sight integral for $\hat{\mathscr{P}}_{ab}^{[M]}$ with $M<N$ \cite{2ndBoltzmannPitrou}. In principle, we can extend the calculation to any orders we want iteratively, limited only by computational power and human frailty.

\subsection{Weak Lensing at 2nd order}\label{Sec2ndLensing}
In this section, we derive the weak lensing effect on the CMB power spectrum from the 2nd-order Boltzmann equation. We will match our result with the previous approach using the remapping approach, i.e. \eqref{mappingapproach}, explicitly stating the assumptions needed for the matching, and explaining their physical significance. We will generalize the derivation to arbitrarily high orders in the next subsection.

First of all, we expand the source term \eqref{sourcetermQ} to 2nd order 
\begin{eqnarray}\label{2ndSourceTermQ}
 \mathcal{Q}_{ab}^{[\text{II}]}=
 \mathfrak{D}_{ab}^{[\text{II}]}
 -2\left(\frac{\text{d}p^0}{\text{d}\eta}\right)^{[\text{I}]}\frac{\partial\mathscr{P}_{ab}^{[\text{I}]}}{\partial p^0}
 -\left(\frac{\text{d}p^0}{\text{d}\eta}\right)^{[\text{II}]}\frac{\partial\bar{\mathscr{P}}_{ab}}{\partial p^0}
 -2\left(\frac{\text{d}x^I}{\text{d}\eta}\right)^{[\text{I}]}\frac{\partial\mathscr{P}_{ab}^{[\text{I}]}}{\partial x^I}
 -2\left(\frac{\text{d}n^i}{\text{d}\eta}\right)^{[\text{I}]}\frac{\partial\mathscr{P}_{ab}^{[\text{I}]}}{\partial n^i}.~~~~~
\end{eqnarray}
The first term on the R.H.S.~in \eqref{2ndSourceTermQ} corresponds to Compton scattering, the second and third terms are related to the redshifts due to the perturbed metric along the light path, the fourth term is responsible to the time-delay effect, and the last term is the weak lensing effect.

Dropping all the other terms except the weak lensing term in \eqref{dxdetaI}, and after performing Fourier transformation and energy integration, we have\footnote{We omit the superscript [I] denoting the 1st-order perturbations and replace $k_{1,J}={_\xi\bar{e}^j}_J k_{1,j}$ with $k_{1,j}$ for simplicity.}
\begin{eqnarray}\label{LensingMatrix}
 \mathcal{\hat{Q}}_{ab}^{[\text{II}]}(\eta,\mathbf{k},\hat{\mathbf{n}})&=&\int \frac{\text{d}\mathbf{k}_1\text{d}\mathbf{k}_2}{(2\pi)^{\frac{3}{2}}}\delta(\mathbf{k}-\mathbf{k}_1-\mathbf{k}_2)2iS^{ij}k_{1,j}\left[\Phi(\eta,\mathbf{k}_1)+\Psi(\eta,\mathbf{k}_1)\right]\frac{\partial}{\partial n^i}\hat{\mathscr{P}}_{ab}(\eta,\mathbf{k}_2,\hat{\mathbf{n}}),
\end{eqnarray}
where $\delta(\mathbf{k})$ is the Dirac delta function. From \eqref{LensingMatrix}, we can see that we have to elaborate $\hat{\mathscr{P}}_{ab}(\eta,\mathbf{k}_2,\hat{\mathbf{n}})$ to 1st order. From now on, we focus on the CMB temperature anisotropies
\begin{equation} 
  \Theta^{[\text{I}]}(\eta,\mathbf{k},\hat{\mathbf{n}})\equiv\frac{1}{4}\hat{I}^{[\text{I}]}(\eta,\mathbf{k},\hat{\mathbf{n}}) 
\end{equation}
only,  though the formalism can be generalized to polarization easily.

With the line of sight approach as shown in \cite{LineOfSightSeljak}, we find that\footnote{We replace $\Delta$ used in \cite{LineOfSightSeljak} with $\Theta$.}
\begin{eqnarray}\label{1stDelta}
 \Theta^{\text{[I]}}(\eta,\mathbf{k},\hat{\mathbf{n}})&=&e^{\bar{\tau}(\eta)}\int^\eta_0 \text{d}\tilde{\eta} e^{ik\mu(\tilde{\eta}-\eta)-\bar{\tau}(\tilde{\eta})}\left\{\Psi'-ik\mu\Phi+\bar{\tau}'\left[\Theta_0-i\mu v_b+\frac{1}{2}(1-3\mu^2)\Pi\right]\right\}~~~\nonumber\\
 &\equiv&e^{\bar{\tau}(\eta)}\int^\eta_0 \text{d}\tilde{\eta} e^{ik\mu(\tilde{\eta}-\eta)}\tilde{S}_T(\tilde{\eta},\mathbf{k},\hat{\mathbf{n}}),
\end{eqnarray}
where the superscript primes denote derivatives with respect to $\eta$, $\mu\equiv \mathbf{k}\cdot\hat{\mathbf{n}}/k$, $v_b$ is the velocity of baryons, $\bar{\tau}(\eta)$ is the background-order optical depth, $\Theta_0$ is the monopole of the temperature anisotropies and $\Pi\equiv(\hat{I}_2-\sqrt{6}\hat{E}_2)/40$ with the subscript 2 denoting the $\ell=2$ multipoles.

After several integrations by parts, we obtain
\begin{eqnarray}\label{dDeltadn}
 \frac{\partial\Theta^{\text{[I]}}}{\partial n^i}(\eta,\mathbf{k},\hat{\mathbf{n}})=ik_i\left[e^{\bar{\tau}(\eta)}\int^\eta_0\text{d}\tilde{\eta} e^{ik\mu(\tilde{\eta}-\eta)}(\tilde{\eta}-\eta)S_T(\tilde{\eta},\mathbf{k})+\frac{3}{2k^2}\bar{\tau}'(\eta)\Pi(\eta,\mathbf{k})\right],
\end{eqnarray}
where the source term
\begin{eqnarray}\label{1stSourceTerm}
S_T\equiv e^{-\bar{\tau}}(\Phi'+\Psi')+g\left(\Theta_0+\Phi+\frac{v_b'}{k}+\frac{\Pi}{2}+\frac{3}{2k^2}\Pi''\right)+g'\left(\frac{v_b}{k}+\frac{3}{k^2}\Pi'\right)+g''\left(\frac{3}{2k^2}\Pi\right)~~~
\end{eqnarray}
with $g\equiv \bar{\tau}' e^{-\bar{\tau}}$. 

To solve the 2nd-order lensed temperature anisotropies $\tilde{\Theta}^{[\text{II}]}(\hat{\mathbf{n}})$, we substitute \eqref{LensingMatrix} into \eqref{PabNow} and replace $\hat{\mathscr{P}}_{ab}$ with its 1st-order scalar component $\Theta^\text{[I]}$. Then, by using \eqref{dDeltadn}, the lensed temperature anisotropies can be written as
\begin{eqnarray}\label{2ndLensingTemp}
 \tilde{\Theta}^{[\text{II}]}(\hat{\mathbf{n}})=\int\frac{\text{d}\mathbf{k_1}\text{d}\mathbf{k_2}}{(2\pi)^3}\int^{\eta_0}_0\text{d}r~2 k_1 k_2 (\mu-\mu_1 \mu_2)[\Phi(\eta,\mathbf{k}_1) + \Psi(\eta,\mathbf{k}_1)]e^{-ik_1\mu_1 r}\nonumber\\
 \left[\int^{\eta_0}_r \text{d}\tilde{r}e^{-ik_2 \mu_2 \tilde{r}}(\tilde{r}-r)S_T(\tilde{\eta},\mathbf{k}_2)-\frac{3}{2k_2^2}e^{-ik_2 \mu_2 r}g(\eta)\Pi(\eta,\mathbf{k}_2)\right],
\end{eqnarray}
where 
\begin{equation}
\mu\equiv\hat{\mathbf{k}}_1\cdot\hat{\mathbf{k}}_2~,~\mu_1\equiv\hat{\mathbf{n}}\cdot\hat{\mathbf{k}}_1~,~\mu_2\equiv\hat{\mathbf{n}}\cdot\hat{\mathbf{k}}_2~,~r\equiv\eta_0-\eta~,~\tilde{r}\equiv\eta_0-\tilde{\eta}.
\end{equation}

As it is well known, the ``pivot'' factor $(\tilde{r}-r)$ in \eqref{2ndLensingTemp} is crucial for boosting the contribution of lensing effect. Even though the leading order of lensing effect is in 2nd order, it has significant influence on the CMB power spectrum which is mainly contributed by linear perturbations. The reason is the following. Most contributions of the source term $S_T$ come from the LSS, i.e.~$\tilde{r}\approx r_{\text{LSS}}$ in \eqref{2ndLensingTemp}. With the lenses at redshift $z<10$, the factor $(\tilde{r}-r)$ is of order $10^4$ and thus compensates for the suppression from the factor of $\sim 10^{-5}$ due to an extra order of perturbation.

We will now re-derive the remapping formula \eqref{mappingapproach} and \eqref{mappingequation} of the weak lensing from the Boltzmann equation. Unlike the familiar approach \cite{LensingFullLewis} where the lensing deflection is calculated via solving the geodesic equation given the metric perturbations, we \emph{explicitly solve the 2nd-order Boltzmann equation using a Green's function method}. This means that we have full control over the entire evolution of the photon distribution from the LSS to today, allowing us to make clear the exact approximations required to recover \eqref{mappingapproach} and \eqref{mappingequation} as follows:
\begin{enumerate}
 \item\label{ApproPi}
  We neglect the term with $\Pi$ in \eqref{2ndLensingTemp} and re-write the equation as
  \begin{eqnarray}\label{Approx1Eq}
    \tilde{\Theta}^{[\text{II}]}(\hat{\mathbf{n}})&=&-2 S^{ij}\int^{\eta_0}_0\text{d}r\frac{\partial}{\partial n^i}\left\{\int\frac{\text{d}\mathbf{k}_1 }{(2\pi)^{\frac{3}{2}}}e^{-i\mathbf{k}_1\cdot\hat{\mathbf{n}}r}\left[\Phi(\eta,\mathbf{k}_1)+\Psi(\eta,\mathbf{k}_1)\right]\right\}\nonumber\\
    &~&~~~~~~~~~~~\int^{\eta_0}_r \text{d}\tilde{r}\frac{\tilde{r}-r}{r\tilde{r}}\frac{\partial}{\partial n^j}\left[\int\frac{d\mathbf{k}_2}{(2\pi)^{\frac{3}{2}}}e^{-i\mathbf{k}_2\cdot\hat{\mathbf{n}}\tilde{r}}S_T(\tilde{\eta},\mathbf{k}_2)\right].
  \end{eqnarray}
  From the integration limits of the double integral, we can see that the lenses at $r$ can only distort the signals of sources further away, i.e.~from $r$ to $\eta_0$, as expected. The neglected term here comes from the boundary condition when we perform the integration by parts in \eqref{dDeltadn}. In general, these boundary terms appear only for those terms with modes $\ell\ge 2$ in the source function $\tilde{S}_T$.
 \item\label{SingleSourceAppro2}
  Then, we replace $(\tilde{r}-r)/\tilde{r}$ with $(r_{\text{LSS}}-r)/r_{\text{LSS}}$, contract $\int^{\eta_0}_0\text{d}r$ to $\int^{r_{\text{LSS}}}_0\text{d}r$ and extend $\int^{\eta_0}_r\text{d}\tilde{r}$ to $\int^{\eta_0}_0\text{d}\tilde{r}$ (i.e.~effectively uncoupling the two integrals). We get
  \begin{equation} \label{Approx1EqB}
  \tilde{\Theta}^{[\text{II}]}(\hat{\mathbf{n}})=2 S^{ij}\frac{\partial}{\partial n^i}\psi(\hat{\mathbf{n}}) \frac{\partial}{\partial n^j}\Theta(\hat{\mathbf{n}})=2 \nabla_{\hat{\mathbf{n}}}\psi(\hat{\mathbf{n}})\cdot \nabla_{\hat{\mathbf{n}}}\Theta(\hat{\mathbf{n}}),
  \end{equation} 
  where $\psi(\hat{\mathbf{n}})$ is the lensing potential defined in \eqref{mappingequation} and we have used the fact that the screen-projected directional derivative
  \begin{equation}\label{NablaSij}
  S^{ij}\frac{\partial}{\partial n^j}=(\nabla_{\hat{\mathbf{n}}})^i
  \end{equation}
in the second equality. This approximation implies that we first integrate all the unlensed CMB signals -- including late-time effects, such as ISW effect and Compton scattering at reionization --  and then we treat those unlensed signals as a single source at the LSS and distort them by lenses between the LSS and the observer.
\end{enumerate}

If we do not perform the integration by parts to replace $\tilde{S}_T(\eta,\mathbf{k},\hat{\mathbf{n}})$ with $S_T(\eta,\mathbf{k})$, the first approximation will not be needed\footnote{See the generalized derivation for high orders in Section \ref{HighOrderLensing}.}. In other words, the first approximation can be folded into the second approximation. Here, we separate them in order to compare our result to previous studies. This ``single-source approximation'' has been evaluated in \cite{LensCorrectionFixedSc} for the weak lensing effect on the CMB power spectra. At 2nd order, their equation (Eqn.~(7)) for redshift-varying sources is exactly the same as \eqref{Approx1Eq}. However, the term with $\Pi$ is missing in their study. In principle, we have to take into account the missing boundary terms when we assume the single-source approximation.

\subsection{Weak Lensing at High Orders}\label{HighOrderLensing}
We will now extend the calculation of lensing effect to arbitrarily high orders and complete the formalism. At 3rd and higher orders, the lensing effect can couple with other effects. For example, the 2nd-order lensed photons can be redshifted by linear metric perturbations given by the term
\begin{equation}
 \left(\frac{\text{d}p^0}{\text{d}\eta}\right)^{[\text{I}]}\frac{\partial \mathscr{P}_{ab}^{[\text{II}]}}{\partial p^0}.
\end{equation}
Thus, it can be ambiguous to distinguish the lensing effect from other effects in high orders. 

We will demonstrate how to derive the usual remapping approach \cite{LensingFullLewis} by explicitly solving the Boltzmann equation. Throughout the derivation, we will clarify all the assumptions needed, and then  validate these approximations in Section \ref{SecDiscussion}.

We start with the source term \eqref{sourcetermQ} and make the following approximations:
\begin{enumerate}
 \item\label{ApproCol} \emph{Ignore non-linear collision terms}: We include only the 1st-order $\mathfrak{D}_{ab}^{[\text{I}]}$ from Compton scattering in \eqref{CollisionTerm}. Nevertheless, we remark that some non-linear collision-related effects can be important, such as the SZ effect on small scales at late time. However, we assume that the lensing effects on them are small and thus these effects can be studied separately\footnote{See \cite{ISWLensingSpergel,SZLensingCooray} for the SZ effect on bispectrum.}. Without the non-linear collision terms, there are no distortions on the frequency spectra of $\mathscr{P}_{ab}$.
 \item\label{ApproTimeDelay}\emph{No time-delay}: We drop all the perturbations on $\text{d}x^I/\text{d}\eta$. This means that when we accumulate the lensing effects, we perform the time integration along straight lines in the spacetime. This term is responsible for the Born approximation and the time-delay effect. 
 \item\label{ApproRedshift}\emph{No redshifting by metric perturbations at non-linear orders}: Similarly, we drop all the perturbations on $\text{d}p^0/\text{d}\eta$ except 
 \begin{equation}
 \left(\frac{\text{d}p^0}{\text{d}\eta}\right)^{[\text{I}]}\frac{\partial \bar{\mathscr{P}}_{ab}}{\partial p^0},
 \end{equation}
 which is responsible to the Sachs-Wolfe and ISW effects in the linear order. In other words, we ignore any contributions due to redshifting beyond the linear order. In principle, redshift-related late-time effects, such as the Rees-Sciama (RS) effect, may be important\footnote{See \cite{RSPowerSpectrum,RSBispectrum} for the RS effect on the CMB power spectrum and bispectrum respectively.}. In Section \ref{SecDiscussion}, we will see that the high-order couplings between the lensing effect and the late-time redshifting are subdominant compared to the pure lensing effect.
 \item\label{ApproNonlinear} \emph{Ignore cross terms between metric perturbations}: We approximate the term $\text{d}n^i/\text{d}\eta$ as
 \begin{equation}\label{NewtonianApprox}
  \frac{\text{d}n^i}{\text{d}\eta}=\sum_{N=1}^{\infty}\frac{1}{N!}\left(\frac{\text{d}n^i}{\text{d}\eta}\right)^{[N]}\approx -\sum_{N=1}^{\infty}\frac{S^{ij}}{N!}\partial_J(\Psi^{[N]}+\Phi^{[N]})\equiv -S^{ij}\partial_J(\Psi^{\text{NL}}+\Phi^{\text{NL}}).
 \end{equation}
 At $N$th order, it means that we ignore all the cross terms of lower-order (less than $N$) perturbations. We call this the \emph{Newtonian approximation} because we drop the cross terms and linearize the General Relativity (GR) as if in Newtonian gravity. 
\end{enumerate}

With these approximations, the source term $\mathcal{Q}_{ab}$ from \eqref{sourcetermQ} is simplified to 
\begin{eqnarray}\label{BolzmannEqnLensingOnly}
 \mathcal{Q}_{ab}=\mathfrak{D}_{ab}^{[\text{I}]}-\left(\frac{\text{d}p^0}{\text{d}\eta}\right)^{[\text{I}]}\frac{\partial \bar{\mathscr{P}}_{ab}}{\partial p^0}+2S^{ij}\partial_J\Psi_\text{W}^{\text{NL}}\frac{\partial\mathscr{P}_{ab}}{\partial n^i},
\end{eqnarray}
where the non-linear Weyl potential $\Psi_\text{W}^{\text{NL}}\equiv(\Psi^{\text{NL}}+\Phi^{\text{NL}})/2$. Physically, the first three approximations imply that we consider only the pure lensing effects acting on the 1st-order intensity matrix. In principle, we have to consider the non-linear intensity matrix, especially when we calculate the $N$-point correlation at large $N$. Having said that, we expect that the lensing effects on the 2nd-order (and higher order) intensity matrix generated at recombination are negligible in the CMB temperature power spectrum and bispectrum. For the power spectrum, it is because $|\Theta^\text{[I]}|\gg|\Theta^\text{[II]}|$. On the other hand, the 2nd-order temperature anisotropies $\Theta^\text{[II]}$ at recombination generate a mild bispectrum \cite{Zhiqi2nd,Su2nd,Fidler2nd}. Thus, the lensing effect on $\Theta^\text{[II]}$ at recombination is expected to be subdominant. However, it may be interesting to study lensing on late-time effects, such as the SZ effect.

We will solve \eqref{BolzmannEqnLensingOnly} order by order in perturbation theory as follows. First, we expand the intensity matrix order by order
\begin{equation}\label{lensedPabSum}
 \hat{\mathscr{P}}_{ab}(\eta_0,\mathbf{k},\hat{\mathbf{n}})=\sum_{N=1}^\infty\frac{1}{N!} \hat{\mathscr{P}}_{ab}^{[N]}(\eta_0,\mathbf{k},\hat{\mathbf{n}}).
\end{equation}
Now we impose the condition, which follows from Approximations \ref{ApproCol} and \ref{ApproRedshift}, that the 1st-order intensity matrix is sourced by the 1st-order collision term $\mathfrak{D}_{ab}^{[\text{I}]}$ and redshifting $\left(\frac{\text{d}p^0}{\text{d}\eta}\right)^{[\text{I}]}$, i.e.~using \eqref{EOMfirstorder}
\begin{equation}
 \frac{\partial\hat{\mathscr{P}}_{ab}^{[\text{I}]}}{\partial\eta}+i\mathbf{k}\cdot\hat{\mathbf{n}} \hat{\mathscr{P}}_{ab}^{[\text{I}]} + \bar{\tau}' \hat{\mathscr{P}}_{ab}^{[\text{I}]} = {S_a}^c {S_b}^d \hat{\mathfrak{D}}_{cd}^{[\text{I}]}+2S_{ab}(\Psi'-i\mathbf{k}\cdot\hat{\mathbf{n}}\Phi).
\end{equation}
On the RHS, we have made use of the fact that there are no polarizations in the background order, i.e.~$\bar{\mathscr{P}}_{ab}=(1/2)\bar{\mathcal{I}}S_{ab}$.
Reinserting the solution $\hat{\mathscr{P}}_{ab}^{[\text{I}]}$ back into \eqref{EOMfirstorder} with Approximations \ref{ApproCol} to \ref{ApproNonlinear}, and omitting for simplicity the $\mathbf{k}$ dependence of $\hat{\mathscr{P}}_{ab}^\text{[II]}$, we get
\begin{equation}
 \frac{\partial\hat{\mathscr{P}}_{ab}^{[\text{II}]}}{\partial\eta}+i\mathbf{k}\cdot\hat{\mathbf{n}} \hat{\mathscr{P}}_{ab}^{[\text{II}]} + \bar{\tau}' \hat{\mathscr{P}}_{ab}^{[\text{II}]} = 2 {S_a}^c {S_b}^d \int\frac{\text{d}\mathbf{k}'\text{d}\mathbf{k}''}{(2\pi)^\frac{3}{2}}
 \delta(\mathbf{k}-\mathbf{k}'-\mathbf{k}'') 2i S^{ij}k'_j\Psi_\text{W}^{\text{NL}}(\mathbf{k}')\frac{\partial\hat{\mathscr{P}}_{cd}^{[\text{I}]}(\mathbf{k}'')}{\partial n^i}.
 \end{equation}
Iterating this, we get at each $N\geq 1$
\begin{equation} \label{eomOrderN}
 \frac{\partial\hat{\mathscr{P}}_{ab}^{[N+1]}}{\partial\eta}+i\mathbf{k}\cdot\hat{\mathbf{n}} \hat{\mathscr{P}}_{ab}^{[N+1]} + \bar{\tau}' \hat{\mathscr{P}}_{ab}^{[N+1]} = \frac{(N+1)!}{N!} {S_a}^c {S_b}^d \int\frac{\text{d}\mathbf{k}'\text{d}\mathbf{k}''}{(2\pi)^\frac{3}{2}}
 \delta(\mathbf{k}-\mathbf{k}'-\mathbf{k}'') 2i S^{ij}k'_j\Psi_\text{W}^{\text{NL}}(\mathbf{k}')\frac{\partial\hat{\mathscr{P}}_{cd}^{[N]}(\mathbf{k}'')}{\partial n^i}.
 \end{equation} 
 
\begin{figure}[ht]
 \begin{tikzpicture}
 \fill (0,0) circle[radius=\dotsize];
 \draw (0,0) node[below]{$\tilde{\eta}$};
 \draw (1,-0.1) node[below]{$<$};
 \fill (2,0) \Square{\dotsize};
 \draw (2,-0.1) node[below]{$\eta_1$};
 \draw (3,-0.1) node[below]{$<$};
 \fill (4,0) \Square{\dotsize};
 \draw (4,-0.1) node[below]{$\eta_2$};
 \draw (5,-0.1) node[below]{$<$};
 \draw (6,-0.1) node[below]{$\hdots$};
 \draw (7,-0.1) node[below]{$<$};
 \fill (8,0) \Square{\dotsize};
 \draw (8,-0.1) node[below]{$\eta_N$};
 \draw (9,-0.1) node[below]{$<$};
 \draw (10,-0.1) node[below]{$\eta_0$};
 \end{tikzpicture}
 \caption{The diagrammatic illustration for the time-ordering of the nested integral in \eqref{fullI_N+1}. The circle node indicates the source while the square nodes indicate the lenses.}
\label{fig:TimeOrdering}
\end{figure}
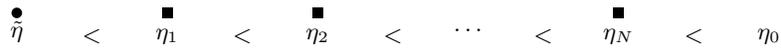

From now on, we focus on the temperature anisotropies. Similar to \eqref{PabNow}, we use the line of sight approach and \eqref{BolzmannEqnLensingOnly} to immediately write down the solution to \eqref{eomOrderN} 
\begin{eqnarray}\label{iterativeI_N+1}
 &~&\frac{1}{(N+1)!}\hat{I}^{[N+1]}(\eta_{N+1},\mathbf{k}_{N+1},\hat{\mathbf{n}})\nonumber\\
 &=& e^{i\mathbf{k}_{N+1}\cdot\hat{\mathbf{n}}r_{N+1}+\bar{\tau}(\eta_{N+1})}
 \int^{\eta_{N+1}}_{0}\text{d}\eta_N\left(-\frac{2}{r_N}\right)\int\frac{\text{d}\mathbf{k}'_N\text{d}\mathbf{k}_N}{(2\pi)^\frac{3}{2}}
 \delta(\mathbf{k}_{N+1}-\mathbf{k}'_{N}-\mathbf{k}_{N})~~\nonumber\\
 &~&\nabla_{\hat{\mathbf{n}}}^{i_N}\left[e^{-i\mathbf{k}'_N\cdot\hat{\mathbf{n}}r_N}\Psi^\text{NL}_\text{W}(\eta_N,\mathbf{k}'_N)\right]\left(\frac{\partial}{\partial n^{i_N}}+ik_{N,i_N}r_N\right)\left[e^{-i\mathbf{k}_N\cdot\hat{\mathbf{n}}r_N-\bar{\tau}(\eta_N)}
 \frac{1}{N!}\hat{I}^{[N]}(\eta_N,\mathbf{k}_N,\hat{\mathbf{n}})\right],
\end{eqnarray}
where $N\geq 1$ denotes the $N$th time of iteration, $r_N\equiv\eta_0-\eta_{N}$ and $\hat{I}^{[1]}(\eta_1,\mathbf{k}_1,\hat{\mathbf{n}})$ is the 1st-order photon intensity, i.e.~$4\Theta^{[\text{I}]}(\eta_1,\mathbf{k}_1,\hat{\mathbf{n}})$ in \eqref{1stDelta}. For $\eta_{N+1}=\eta_0$ at present, we can write the $(N+1)$th iteration explicitly as the following $N$-nested integral
\begin{eqnarray}\label{fullI_N+1}
 &~&\frac{1}{(N+1)!}\hat{I}^{[N+1]}(\eta_0,\mathbf{k}_{N+1},\hat{\mathbf{n}})\nonumber\\
 &=&\int^{\eta_0}_0\text{d}\tilde{\eta}~\Bigg\{
 \int^{\eta_{0}}_{\tilde{\eta}}\text{d}\eta_N\left(-\frac{2}{r_N}\right)\int\frac{\text{d}\mathbf{k}'_N\text{d}\mathbf{k}_N}{(2\pi)^\frac{3}{2}}
 \delta(\mathbf{k}_{N+1}-\mathbf{k}'_{N}-\mathbf{k}_{N})
 \nabla_{\hat{\mathbf{n}}}^{i_N}\left[e^{-i\mathbf{k}'_N\cdot\hat{\mathbf{n}}r_N}\Psi^\text{NL}_\text{W}(\eta_N,\mathbf{k}'_N)\right]\left(\frac{\partial}{\partial n^{i_N}}+ik_{N,i_N}r_N\right)\bigg\{\nonumber\\
 &~&~~~~~~~~~~~~~~~~~~~~~~~~~~~~~~~~~~~~~~~~~~~~~~~~~~~~~~~~~~~~~~~~~~~~~~
 \vdots\nonumber\\
 &~&~~~~~~~~~~~~~~~~~
 \int^{\eta_{3}}_{\tilde{\eta}}\text{d}\eta_2\left(-\frac{2}{r_2}\right)\int\frac{\text{d}\mathbf{k}'_2\text{d}\mathbf{k}_2}{(2\pi)^\frac{3}{2}}
 \delta(\mathbf{k}_{3}-\mathbf{k}'_{2}-\mathbf{k}_{2})
 \nabla_{\hat{\mathbf{n}}}^{i_2}\left[e^{-i\mathbf{k}'_2\cdot\hat{\mathbf{n}}r_2}\Psi^\text{NL}_\text{W}(\eta_2,\mathbf{k}'_2)\right]\left(\frac{\partial}{\partial n^{i_2}}+ik_{2,i_2}r_2\right)\bigg\{\nonumber\\
 &~&~~~~~~~~~~~~~~~~~~~~~~
 \int^{\eta_{2}}_{\tilde{\eta}}\text{d}\eta_1\left(-\frac{2}{r_1}\right)\int\frac{\text{d}\mathbf{k}'_1\text{d}\mathbf{k}_1}{(2\pi)^\frac{3}{2}}
 \delta(\mathbf{k}_{2}-\mathbf{k}'_{1}-\mathbf{k}_{1})
 \nabla_{\hat{\mathbf{n}}}^{i_1}\left[e^{-i\mathbf{k}'_1\cdot\hat{\mathbf{n}}r_1}\Psi^\text{NL}_\text{W}(\eta_1,\mathbf{k}'_1)\right]\left(\frac{\partial}{\partial n^{i_1}}+ik_{1,i_1}r_1\right)\bigg\{\nonumber\\
 &~&~~~~~~~~~~~~~~~~~~~~~~~~~~~~~~~~~~~~~~~~~~~~~~~~~~~~~~~
 e^{-i\mathbf{k}_1\cdot\hat{\mathbf{n}}\tilde{r}}4\tilde{S}_T(\tilde{\eta},\mathbf{k}_1,\hat{\mathbf{n}}) 
 ~~~~~\bigg\}\bigg\}\hdots\bigg\}\Bigg\}.
\end{eqnarray}
In \eqref{fullI_N+1}, we have used a familiar stratagem often encountered in quantum field theory to rearrange the order of the integration limits
\begin{eqnarray}
 \int^{\eta_{M+1}}_0\text{d}\eta_M\int^{\eta_M}_0\text{d}\tilde{\eta}=\int^{\eta_{M+1}}_0\text{d}\tilde{\eta}\int^{\eta_{M+1}}_{\tilde{\eta}}\text{d}\eta_M
\end{eqnarray} 
to pull out the innermost integration with respect to $\tilde{\eta}$ which integrates over the sources. This trick allows us to explicitly expose the physical meaning of the nested integral -- the lens at $\eta_N$, which generates the $(N+1)$th-order intensity from the $N$th-order intensity, must be located \emph{after} the lenses at $\eta_{M}$ ($M<N$) but \emph{before} the observer at $\eta_0$. The time-ordering is illustrated in Fig.~\ref{fig:TimeOrdering}.

Following \eqref{lensedPabSum}, the lensed photon intensity at present is the sum of all possible iterations, i.e.~
\begin{eqnarray}\label{lensedISum}
 \hat{I}(\eta_0,\mathbf{k},\hat{\mathbf{n}})=\sum_{N=1}^\infty \frac{1}{N!}\hat{I}^{[N]}(\eta_0,\mathbf{k},\hat{\mathbf{n}}).
\end{eqnarray}
\eqref{fullI_N+1} and \eqref{lensedISum} form the complete formula for the pure lensing effect on the 1st-order photon intensity and include all possible ways to distort the 1st-order photon intensity with the weak lensing effect. \\

In the following, we proceed to recover the remapping approach of the lensing effect in \eqref{mappingapproach} and \eqref{mappingequation} from the formulae shown in \eqref{fullI_N+1} and \eqref{lensedISum}. Several further approximations have to be made and we will clarify them explicitly. Before we do that, we first commute the terms $\partial/\partial n^i+ik_i r$ with the terms $\nabla_{\hat{\mathbf{n}}}^i[e^{-i\mathbf{k}'\cdot\hat{\mathbf{n}}r}\Psi^{\text{NL}}_{\text{W}}(\eta,\mathbf{k}')]$ and rearrange the former to the right hand side of the latter in \eqref{fullI_N+1}. The commutation can be achieved by using\footnote{We omit the dependences of $\Psi^{\text{NL}}_{\text{W}}$ on the R.H.S. of \eqref{NablaCommutation}.}
\begin{eqnarray}\label{NablaCommutation}
 &~&(\frac{\partial}{\partial n^{i_N}}+ik_{N,i_N}r_N)\left\{\nabla_{\hat{\mathbf{n}}}^{i_M}\left[e^{-i\mathbf{k}'_M\cdot\hat{\mathbf{n}}r_M}\Psi^{\text{NL}}_{\text{W}}(\eta_M,\mathbf{k}'_M)\right]F(\cdot)\right\}\nonumber\\
 &=&\frac{\partial S^{i_Mj_M}}{\partial n^{i_N}}\frac{\partial}{\partial n^{j_M}}\left[e^{-i\mathbf{k}'_M\cdot\hat{\mathbf{n}}r_M}\Psi^{\text{NL}}_{\text{W}}\right]F(\cdot)+\nabla_{\hat{\mathbf{n}}}^{i_M}\left[e^{-i\mathbf{k}'_M\cdot\hat{\mathbf{n}}r_M}\Psi^{\text{NL}}_{\text{W}}\right]\left(\frac{\partial}{\partial n^{i_N}}-ik'_{M,i_N}r_M+ik_{N,i_N}r_N\right)F(\cdot)
\end{eqnarray}
iteratively for $M<N$, with $F(\cdot)$ as some arbitrary function and we have used \eqref{NablaSij} to commute the two derivatives
\begin{equation}
\frac{\partial}{\partial n^{i_{N}}} (\nabla_{\hat{\mathbf{n}}})^{i_{M}}= \frac{\partial}{\partial n^{i_{N}}} \left(S^{{i_{M}}{j_{M}}}\frac{\partial}{\partial n^{j_{M}}} \right)=\frac{\partial S^{{i_{M}}{j_{M}}} }{\partial n^{i_{N}}} \frac{\partial}{\partial n^{j_{M}}} + S^{{i_{M}}{j_{M}}}\frac{\partial}{\partial n^{i_{N}}}\frac{\partial}{\partial n^{j_{M}}}.
\end{equation}
Now, we list out the extra approximations needed:
\begin{enumerate}
 \setcounter{enumi}{4}
 \item\label{ApprodSdn}
\emph{Neglect the directional derivative of the screen projector}. This means that 
\begin{equation}\label{dSijdn=0}
 \frac{\partial S^{i_Mj_M}}{\partial n^{i_N}}\approx 0.
\end{equation}

We can then move all the partial derivatives into the innermost integrand to act on $\tilde{S}_T(\tilde{\eta}$), and rewrite \eqref{fullI_N+1} as 
 \begin{eqnarray}\label{iterativeI_NCommutedI}
  &~&\frac{1}{(N+1)!}\hat{I}^{[N+1]}(\eta_0,\mathbf{k}_{N+1},\hat{\mathbf{n}})\nonumber\\
  &=&\int^{\eta_0}_0\text{d}\tilde{\eta}~\Bigg\{
  \int^{\eta_{0}}_{\tilde{\eta}}\text{d}\eta_N\left(-\frac{2}{r_N}\right)\int\frac{\text{d}\mathbf{k}'_N\text{d}\mathbf{k}_N}{(2\pi)^\frac{3}{2}}
  \delta(\mathbf{k}_{N+1}-\mathbf{k}'_{N}-\mathbf{k}_{N})
  \nabla_{\hat{\mathbf{n}}}^{i_N}\left[e^{-i\mathbf{k}'_N\cdot\hat{\mathbf{n}}r_N}\Psi^\text{NL}_\text{W}(\eta_N,\mathbf{k}'_N)\right]\bigg\{\nonumber\\
  &~&~~~~~~~~~~~~~~~~~~~~~~~~~~~~~~~~~~~~~~~~~~~~~~~~~~~~~~~~~~~~\vdots\nonumber\\
  &~&~~~~~~~~~~~~~~~~~
  \int^{\eta_{3}}_{\tilde{\eta}}\text{d}\eta_2\left(-\frac{2}{r_2}\right)\int\frac{\text{d}\mathbf{k}'_2\text{d}\mathbf{k}_2}{(2\pi)^\frac{3}{2}}
  \delta(\mathbf{k}_{3}-\mathbf{k}'_{2}-\mathbf{k}_{2})
  \nabla_{\hat{\mathbf{n}}}^{i_2}\left[e^{-i\mathbf{k}'_2\cdot\hat{\mathbf{n}}r_2}\Psi^\text{NL}_\text{W}(\eta_2,\mathbf{k}'_2)\right]\bigg\{\nonumber\\
  &~&~~~~~~~~~~~~~~~~~~~~~~
  \int^{\eta_{2}}_{\tilde{\eta}}\text{d}\eta_1\left(-\frac{2}{r_1}\right)\int\frac{\text{d}\mathbf{k}'_1\text{d}\mathbf{k}_1}{(2\pi)^\frac{3}{2}}
  \delta(\mathbf{k}_{2}-\mathbf{k}'_{1}-\mathbf{k}_{1})
  \nabla_{\hat{\mathbf{n}}}^{i_1}\left[e^{-i\mathbf{k}'_1\cdot\hat{\mathbf{n}}r_1}\Psi^\text{NL}_\text{W}(\eta_1,\mathbf{k}'_1)\right]\bigg\{\nonumber\\
  &~&\bigg[\frac{\partial}{\partial n^{i_N}}+i\sum^{N-1}_{M=1}k'_{M,i_N}(r_N-r_M)+ik_{1,i_N}r_N\bigg]\hdots
  \bigg[\frac{\partial}{\partial n^{i_2}}+ik'_{1,i_2}(r_2-r_1)+ik_{1,i_2}r_2\bigg]
  \bigg[\frac{\partial}{\partial n^{i_1}}+ik_{1,i_1}r_1\bigg]\nonumber\\
  &~&~~~~~~~~~~~~~~~~~~~~~~~~~~~~~~~~~~~~~~~~~~~~~
  e^{-i\mathbf{k}_1\cdot\hat{\mathbf{n}}\tilde{r}}4\tilde{S}_T(\tilde{\eta},\mathbf{k}_1,\hat{\mathbf{n}})~~~~~~~~~~\bigg\}\bigg\}\hdots\bigg\}\Bigg\}.
 \end{eqnarray}
 This approximation is related to the fact that the argument $(\hat{\mathbf{n}}+\boldsymbol{\alpha})$ in \eqref{mappingapproach} is not a unit vector. Although this approximation is implied in the remapping approach, it has not been clarified in literature -- this demonstrates the value of our full Boltzmann equation approach. Nevertheless, the leading correction from this approximation is at 3rd order\footnote{The directional derivative of the screen projector requires $\partial/\partial n^i$ from a lens acting on $\nabla^j_{\hat{\mathbf{n}}}$ of another lens as shown in \eqref{fullI_N+1}. Therefore, the photon intensity is at least in 3rd order including the 1st-order source term $\tilde{S}_T$.} and thus we do not expect significant corrections on the CMB power spectra and bispectra.

 In Section \ref{DiagramRep}, we will develop a set of diagrams to represent different couplings in \eqref{iterativeI_NCommutedI}. In addition, we show that \eqref{iterativeI_NCommutedI} can formally be expressed as a Dyson series in Appendix \ref{AppendixA}.
 \item\label{ApproNolenslens} \emph{Ignore lens-lens couplings}. Operationally, this means that we drop all the terms with $k'_M$ for any $M$ in the second last line of \eqref{iterativeI_NCommutedI}. Physically, we ignore all the lensing effects on lenses, i.e.~lens-lens couplings \cite{2ndBornWHu}. We will consider the effects of the lens-lens couplings in Section \ref{ValidApprox}, but for the moment we continue our re-derivation of the remapping approach.  This approximation decouples all the Fourier integrals, allowing us to immediately write down the photon intensity in configuration space as
 \begin{eqnarray}\label{NoLensCouplingsI}
  &~&\frac{1}{(N+1)!}\hat{I}^{[N+1]}(\eta_0,\hat{\mathbf{n}})\nonumber\\
  &=&\int^{\eta_0}_0\text{d}\tilde{\eta}~\Bigg\{
  \int^{\eta_{0}}_{\tilde{\eta}}\text{d}\eta_N\left(-\frac{2}{r_N}\right)
  \nabla_{\hat{\mathbf{n}}}^{i_N}\left[\Psi^\text{NL}_\text{W}(\eta_N,-\hat{\mathbf{n}}r_N)\right]\bigg\{\hdots\nonumber\\
  &~&~~~~~~~~~~~~~~~~~
  \int^{\eta_{3}}_{\tilde{\eta}}\text{d}\eta_2\left(-\frac{2}{r_2}\right)
  \nabla_{\hat{\mathbf{n}}}^{i_2}\left[\Psi^\text{NL}_\text{W}(\eta_2,-\hat{\mathbf{n}}r_2)\right]\bigg\{\nonumber\\
  &~&~~~~~~~~~~~~~~~~~~~~~~
  \int^{\eta_{2}}_{\tilde{\eta}}\text{d}\eta_1\left(-\frac{2}{r_1}\right)
  \nabla_{\hat{\mathbf{n}}}^{i_1}\left[\Psi^\text{NL}_\text{W}(\eta_1,-\hat{\mathbf{n}}r_1)\right]\bigg\{\nonumber\\
  &~&\int\frac{\text{d}\mathbf{k}_1}{(2\pi)^\frac{3}{2}}
  \left(\frac{\partial}{\partial n^{i_N}}+ik_{1,i_N}r_N\right)\hdots
  \left(\frac{\partial}{\partial n^{i_2}}+ik_{1,i_2}r_2\right)
  \left(\frac{\partial}{\partial n^{i_1}}+ik_{1,i_1}r_1\right)\nonumber\\
  &~&~~~~~~~~~~~~~~~~~~~~~~~e^{-i\mathbf{k}_1\cdot\hat{\mathbf{n}}\tilde{r}}4\tilde{S}_T(\tilde{\eta},\mathbf{k}_1,\hat{\mathbf{n}})\bigg\}\bigg\}\hdots\bigg\}\Bigg\},
 \end{eqnarray}
where 
 \begin{eqnarray}
  \Psi^\text{NL}_\text{W}(\eta_N,-\hat{\mathbf{n}}r_N)=\int\frac{\text{d}\mathbf{k}'_N}{(2\pi)^\frac{3}{2}}e^{-i\mathbf{k}'_N\cdot\hat{\mathbf{n}}r_N}\Psi^\text{NL}_\text{W}(\eta_N,\mathbf{k}'_N).
 \end{eqnarray}
 \item\label{SingleSourceApproH} \emph{Place all unlensed CMB signals at the LSS.} Explicitly, we replace
\begin{eqnarray}
\int^{\eta_{M+1}}_{\tilde{\eta}}\text{d}\eta_M &\longrightarrow& \int^{\eta_{M+1}}_{\eta_\text{LSS}}\text{d}\eta_M, \\
\left[\frac{\partial}{\partial n^{i_M}} + ik_{1,i_M}r_M\right]e^{-i\mathbf{k}_1\cdot\hat{\mathbf{n}}\tilde{r}}&=&
 -\frac{r_M-\tilde{r}}{\tilde{r}}\frac{\partial}{\partial n^{i_M}}e^{-i\mathbf{k}_1\cdot\hat{\mathbf{n}}\tilde{r}}\longrightarrow -\frac{r_M-r_{\text{LSS}}}{r_{\text{LSS}}}\frac{\partial}{\partial n^{i_M}}e^{-i\mathbf{k}_1\cdot\hat{\mathbf{n}}\tilde{r}}
\end{eqnarray}
for any $M$. Notice that in order  to recover the remapping approach, we \emph{do not} replace $\tilde{r}$ with $r_\text{LSS}$ in the exponential. It means that when we do the line of sight approach for the unlensed temperature anisotropies, we consider the effects at various times properly. In contrast, when we calculate the lensing effects on the source term, we treat the sources as if they are all located at the LSS.
  
  This approximation decouples all the time integrations such that the nested integral is deconvolved into a product of integrals as follows
 \begin{eqnarray}\label{SingleSourceI}
  \frac{1}{(N+1)!}\hat{I}^{[N+1]}(\eta_0,\hat{\mathbf{n}})&=&
  2\int^{\eta_{0}}_{\eta_{\text{LSS}}}\text{d}\eta_N\left(\frac{r_N-r_{\text{LSS}}}{r_N~r_{\text{LSS}}}\right)
  \nabla_{\hat{\mathbf{n}}}^{i_N}\left[\Psi^\text{NL}_\text{W}(\eta_N,-\hat{\mathbf{n}}r_N)\right]\bigg\{\hdots\nonumber\\
  &~&~~~~~
  2\int^{\eta_{3}}_{\eta_{\text{LSS}}}\text{d}\eta_2\left(\frac{r_2-r_{\text{LSS}}}{r_2~r_{\text{LSS}}}\right)
  \nabla_{\hat{\mathbf{n}}}^{i_2}\left[\Psi^\text{NL}_\text{W}(\eta_2,-\hat{\mathbf{n}}r_2)\right]\bigg\{\nonumber\\
  &~&~~~~~~~~~~
  2\int^{\eta_{2}}_{\eta_{\text{LSS}}}\text{d}\eta_1\left(\frac{r_1-r_{\text{LSS}}}{r_1~r_{\text{LSS}}}\right)
  \nabla_{\hat{\mathbf{n}}}^{i_1}\left[\Psi^\text{NL}_\text{W}(\eta_1,-\hat{\mathbf{n}}r_1)\right]\bigg\{\nonumber\\
  &~&~~~
  \frac{\partial}{\partial n^{i_N}}\hdots \frac{\partial}{\partial n^{i_2}} \frac{\partial}{\partial n^{i_1}}
  \left[\int\frac{\text{d}\mathbf{k}_1}{(2\pi)^\frac{3}{2}}\int^{\eta_0}_0\text{d}\tilde{\eta}
  ~e^{-i\mathbf{k}_1\cdot\hat{\mathbf{n}}\tilde{r}}4S_T(\tilde{\eta},\mathbf{k}_1)\right]~\bigg\}\bigg\}\hdots\bigg\}\nonumber\\
  &=&\frac{4}{N!}\alpha^{i_N}\hdots\alpha^{i_2}\alpha^{i_1}\frac{\partial}{\partial n^{i_N}}\hdots\frac{\partial}{\partial n^{i_2}}\frac{\partial}{\partial n^{i_1}}
  \Theta(\hat{\mathbf{n}}),
 \end{eqnarray}
where the deflection angle $\boldsymbol{\alpha}$ is expressed as
\begin{equation}\label{DeflectionAngle}
 \boldsymbol{\alpha}(\hat{\mathbf{n}}) = 2 \int^{r_{\text{LSS}}}_0 \text{d}r \frac{r-r_{\text{LSS}}}{r~r_{\text{LSS}}}\nabla_{\hat{\mathbf{n}}}\Psi^{\text{NL}}_\text{W}(\eta,-\hat{\mathbf{n}}r)
\end{equation}
and the unlensed temperature anisotropies are
\begin{eqnarray}\label{UnlensedTemp}
  \Theta(\hat{\mathbf{n}}) = \int\frac{\text{d}\mathbf{k}_1}{(2\pi)^\frac{3}{2}}\int^{\eta_0}_0\text{d}\tilde{\eta} ~e^{-  i\mathbf{k}_1\cdot\hat{\mathbf{n}}\tilde{r}}S_T(\tilde{\eta},\mathbf{k}_1).
\end{eqnarray}
\end{enumerate}

Finally, summing up all orders, we recover the remapping approach of the weak lensing effect on the photon intensity \cite{LensingFullLewis}, i.e.~ 
 \begin{eqnarray}\label{OldApproach}
  \hat{I}(\eta_0,\hat{\mathbf{n}})=\sum_{N=1}^\infty \frac{1}{N!}\hat{I}^{[N]}(\eta_0,\hat{\mathbf{n}})=4\Theta(\hat{\mathbf{n}}+\boldsymbol{\alpha})=4\tilde{\Theta}(\hat{\mathbf{n}}),
 \end{eqnarray}
where $\tilde{\Theta}$ denotes the lensed temperature anisotropies.

\section{Power Spectrum from the lens-lens couplings}\label{ValidApprox}
Due to the presence of the boosting factors (i.e.~$(r_N-r_M)$), the lens-lens couplings (Approximation \ref{ApproNolenslens} in Section \ref{HighOrderLensing}) may exhibit non-negligible contributions to the CMB power spectra. This effect has not been explicitly calculated\footnote{Nevertheless, the effects from lens-lens couplings, together with the effects of neglecting the  Born approximation (included in Approximation \ref{ApproTimeDelay}), on background galaxy images have been studied in \cite{2ndBornWHu}.}. In this section, we employ the formalism developed to assess their importance quantitatively on the temperature power spectrum. We will assess the other approximations qualitatively in Section \ref{SecDiscussion}.

We start with \eqref{iterativeI_NCommutedI}, which implies that we applied all the approximations except Approximation \ref{ApproNolenslens} and Approximation \ref{SingleSourceApproH} in Section \ref{HighOrderLensing}. To single out the lens-lens couplings from other approximations, we further apply Approximation \ref{SingleSourceApproH} and express the $(N+1)$th-order intensity as
\begin{eqnarray}
  &~&\frac{1}{(N+1)!}\hat{I}^{[N+1]}(\eta_0,\mathbf{k}_{N+1},\hat{\mathbf{n}})\nonumber\\
  &=&
  \int^{\eta_{0}}_{\eta_\text{LSS}}\text{d}\eta_N\left(-\frac{2}{r_N}\right)\int\frac{\text{d}\mathbf{k}'_N\text{d}\mathbf{k}_N}{(2\pi)^\frac{3}{2}}
  \delta(\mathbf{k}_{N+1}-\mathbf{k}'_{N}-\mathbf{k}_{N})
  \nabla_{\hat{\mathbf{n}}}^{i_N}\left[e^{-i\mathbf{k}'_N\cdot\hat{\mathbf{n}}r_N}\Psi^\text{NL}_\text{W}(\eta_N,\mathbf{k}'_N)\right]\bigg\{\nonumber\\
  &~&~~~~~~~~~~~~~~~~~~~~~~~~~~~~~~~~~~~~~~~~~~~~~~~~~~~~~~~~~~~~\vdots\nonumber\\
  &~&~~~~~~
  \int^{\eta_{3}}_{\eta_\text{LSS}}\text{d}\eta_2\left(-\frac{2}{r_2}\right)\int\frac{\text{d}\mathbf{k}'_2\text{d}\mathbf{k}_2}{(2\pi)^\frac{3}{2}}
  \delta(\mathbf{k}_{3}-\mathbf{k}'_{2}-\mathbf{k}_{2})
  \nabla_{\hat{\mathbf{n}}}^{i_2}\left[e^{-i\mathbf{k}'_2\cdot\hat{\mathbf{n}}r_2}\Psi^\text{NL}_\text{W}(\eta_2,\mathbf{k}'_2)\right]\bigg\{\nonumber\\
  &~&~~~~~~~~~~~~~~
  \int^{\eta_{2}}_{\eta_\text{LSS}}\text{d}\eta_1\left(-\frac{2}{r_1}\right)\int\frac{\text{d}\mathbf{k}'_1\text{d}\mathbf{k}_1}{(2\pi)^\frac{3}{2}}
  \delta(\mathbf{k}_{2}-\mathbf{k}'_{1}-\mathbf{k}_{1})
  \nabla_{\hat{\mathbf{n}}}^{i_1}\left[e^{-i\mathbf{k}'_1\cdot\hat{\mathbf{n}}r_1}\Psi^\text{NL}_\text{W}(\eta_1,\mathbf{k}'_1)\right]\bigg\{\nonumber\\
  &~&\bigg[i\sum^{N-1}_{M=1}k'_{M,i_N}(r_N-r_M)+ik_{1,i_N}(r_N-r_\text{LSS})\bigg]\hdots
  \bigg[ik'_{1,i_2}(r_2-r_1)+ik_{1,i_2}(r_2-r_\text{LSS})\bigg]
  \bigg[ik_{1,i_1}(r_1-r_\text{LSS})\bigg]\nonumber\\
  &~&~~~~~~~~~~~~~~~~~~~~~~~~~~~~~~~~~~~~~~~~~~~~~
  \int^{\eta_0}_0\text{d}\tilde{\eta}~e^{-i\mathbf{k}_1\cdot\hat{\mathbf{n}}\tilde{r}}4S_T(\tilde{\eta},\mathbf{k}_1)~~~~~~~~~~\bigg\}\bigg\}\hdots\bigg\}.
\end{eqnarray}

Applying Approximation 5 (i.e. \eqref{dSijdn=0}), we can express the $(N+1)$th-order lensed temperature anisotropies in configuration space as\footnote{We expand the lensed temperature anisotropies as $\tilde{\Theta}(\hat{\mathbf{n}})=\sum_{N=1}^\infty\tilde{\Theta}^{[N]}(\hat{\mathbf{n}})$.}
\begin{eqnarray}\label{SingleSourceIwithoutApp6}
 \tilde{\Theta}^{[N+1]}(\hat{\mathbf{n}})&=&\frac{1}{4(N+1)!}\hat{I}^{[N+1]}(\eta_0,\hat{\mathbf{n}})\nonumber\\
 &=&
  \int^{\eta_{0}}_{\eta_\text{LSS}}\text{d}\eta_N
  \nabla_{\hat{\mathbf{n}}}^{i_N}\left[\Psi^\text{NL}_\text{W}(\eta_N,-\hat{\mathbf{n}}r_N)\right](\hat{\square}_{\hat{\mathbf{n}},r_N})_{i_N}\bigg\{\nonumber\\
  &~&~~~~~~~~~~~~~~~~~~~~~~~~~~~~~~~~~~~~~~~~~\vdots\nonumber\\
  &~&~~~~~~~~
  \int^{\eta_{3}}_{\eta_\text{LSS}}\text{d}\eta_2
  \nabla_{\hat{\mathbf{n}}}^{i_2}\left[\Psi^\text{NL}_\text{W}(\eta_2,-\hat{\mathbf{n}}r_2)\right](\hat{\square}_{\hat{\mathbf{n}},r_2})_{i_2}\bigg\{ \nonumber\\
  &~&~~~~~~~~~~~~~~~~~~
  \int^{\eta_{2}}_{\eta_\text{LSS}}\text{d}\eta_1
  \nabla_{\hat{\mathbf{n}}}^{i_1}\left[\Psi^\text{NL}_\text{W}(\eta_1,-\hat{\mathbf{n}}r_1)\right](\hat{\square}_{\hat{\mathbf{n}},r_1})_{i_1}~
  \Theta(\hat{\mathbf{n}})~\bigg\}\hdots\bigg\},~~~~~~~~~~~~
\end{eqnarray}
where the unlensed temperature anisotropies $\Theta(\hat{\mathbf{n}})$ are given by \eqref{UnlensedTemp} and the vector operator $\hat{\square}_{\hat{\mathbf{n}},r}$ is defined such that it is non-zero only when it acts on $\mathcal{X}$
\begin{eqnarray}\label{SquareOperator}
 (\hat{\square}_{\hat{\mathbf{n}},r})_i~\mathcal{X}(\eta',-\hat{\mathbf{n}}r')\equiv 2 \frac{r-r'}{r~r'}\frac{\partial}{\partial n^i} \mathcal{X}(\eta',-\hat{\mathbf{n}}r')
\end{eqnarray}
with $\mathcal{X}$ denoting $\Psi^\text{NL}_\text{W}$ or\footnote{With Approximation \ref{SingleSourceApproH}, $r'$ is replaced by $r_\text{LSS}$ when $\hat{\square}_{\hat{\mathbf{n}},r}$ acts on $\Theta$.}
$\Theta$. Physically, acting $\hat{\square}_{\hat{\mathbf{n}},r}$ on $\Psi^\text{NL}_\text{W}$ corresponds to lens-lens couplings while acting on $\Theta$ corresponds to lensing the sources. We will develop a set of diagrams to represent these couplings in the following.

\subsection{Diagrammatic Approach to $N$th-order Correlation Functions} \label{DiagramRep}
In this subsection, we develop a set of rules as a book-keeping tool to compute any $N$th-order correlation functions.  This facilitates the expansion of the lensing effect to higher orders for the calculation of the corresponding CMB power spectra and bispectra. Although we demonstrate the diagrammatic approach based on \eqref{SingleSourceIwithoutApp6}, we note that the diagrams can easily be generalized to include time-varying sources by not applying Approximation \ref{SingleSourceApproH} as in \eqref{DysonI_N+1}. Moreover, similar diagrams can be developed to include the redshift and time-delay effects. In Section \ref{SecDiscussion}, we will argue that these effects are subdominant compared to the lensing effect.

As we can see in \eqref{SingleSourceIwithoutApp6}, the $(N+1)$th-order photon intensity contains time-ordered integrals over two types of interaction terms -- the unlensed temperature anisotropies $\Theta$ (diagrammatically $\begin{tikzpicture}[baseline=(current bounding box.west)] \fill (0,0) circle[radius=\dotsize]; \draw (0,0) node[below]{$0$}; \end{tikzpicture}$ where the $0$ below denotes the ordering)\footnote{Without Approximation \ref{SingleSourceApproH}, the circle node represents $\int^{\eta_0}_0\text{d}\tilde{\eta}~S_T(\tilde{\eta})$ instead of $\Theta$.} and the $N$th potential term $\nabla_{\hat{\mathbf{n}}}[\Psi^\text{NL}_\text{W}(\eta_N)]$ (diagrammatically $\begin{tikzpicture}[baseline=(current bounding box.west)] \fill (0,0) \Square{\dotsize}; \draw (0,0) node[below]{$N$}; \end{tikzpicture}$ where the label $N$ indicates the time-ordering of the potential terms). With $N>M$, it means that the $N$th potential term is located \emph{later} in time compared to the $M$th potential term. We 
write down 
the time-ordered terms from left to right in a row, for example, the 4th-order photon intensity $\hat{I}^{[4]}$ will be  expressed as
\begin{center}
\vskip 2pt
\begin{tikzpicture} \NodesUp{3}{0}{0} \end{tikzpicture}
\vskip 2pt
\end{center}
which is simply a simplified expression for  Fig.~\ref{fig:TimeOrdering}.

In addition to the interaction terms, there is an ``action'' term in \eqref{SquareOperator}, i.e.~the vector operator $\hat{\square}_{\hat{\mathbf{n}},r}$ acting on either interaction term $\Theta$ or $\nabla_{\hat{\mathbf{n}}}[\Psi^\text{NL}_\text{W}]$.  We denote the action by an over-arc, i.e.~
\begin{tikzpicture}[baseline=(current bounding box.west)]
\fill (0,0) circle[radius=\dotsize]; \draw (0,0) node[below]{$0$};
\draw (1,0) node[below]{$\hdots$};
\fill (2,0) \Square{\dotsize}; \draw (2,0) node[below]{$N$};
\draw (2,0) arc (0:180:1);
\end{tikzpicture}
for action of $\hat{\square}_{\hat{\mathbf{n}},r_N}$ on $S_T$ and
\begin{tikzpicture}[baseline=(current bounding box.west)]
\fill (0,0) \Square{\dotsize}; \draw (0,0) node[below]{$M$};
\draw (1,0) node[below]{$\hdots$};
\fill (2,0) \Square{\dotsize}; \draw (2,0) node[below]{$N$};
\draw (2,0) arc (0:180:1);
\end{tikzpicture} 
for action of $\hat{\square}_{\hat{\mathbf{n}},r_N}$ on $\nabla_{\hat{\mathbf{n}}}[\Psi^\text{NL}_\text{W}(\eta_M)]$. For each $N$th action, there is an $N$th time integral $\int_{\eta_\text{LSS}}^{\eta_{N+1}} d\eta_N$ associated with it\footnote{Without Approximation \ref{SingleSourceApproH}, we replace $\eta_\text{LSS}$ by $\tilde{\eta}$. Here, $\tilde{\eta}$ locates the source term $S_T$ and integrates from $0$ to $\eta_0$.}. We summarize these rules in Table \ref{DiagramRuleLen}. 

\begin {table}[htbp]
 \begin{center}
  \begin{tabular}{| p{2.5cm} | p{9.5cm} | p{5.5cm}|}
   \hline ~Diagram& ~Term & ~Physical Meaning\\
   \hline
    ~~~~~~~~~~\begin{tikzpicture}[baseline=(current bounding box.west)] \fill (0,0) circle[radius=\dotsize]; \draw (0,0) node[below]{$0$}; \end{tikzpicture} &
    ~~$\Theta(\hat{\mathbf{n}})$ or $\int^{\eta_0}_0 \text{d}\tilde{\eta}~S_T(\tilde{\eta},-\hat{\mathbf{n}}\tilde{r}) \nonumber$ &
    ~~Unlensed source generated at node 0.\\
   \hline 
    ~~~~~~~~~~\begin{tikzpicture}[baseline=(current bounding box.west)] \fill (0,0) \Square{\dotsize}; \draw (0,0) node[below]{$N$}; \end{tikzpicture} &
    ~~$\nabla_{\hat{\mathbf{n}}}^{i_N}\left[\Psi^\text{NL}_\text{W}(\eta_N,-\hat{\mathbf{n}}r_N)\right] \nonumber$ &
    ~~Lens at node $N$.\\
   \hline ~ & ~ & ~ \\
    \begin{tikzpicture}[baseline=(current bounding box.west)]
     \fill (0,0) circle[radius=\dotsize]; \draw (0,0) node[below]{$0$};
     \draw (1,0) node[below]{$...$};
     \fill (2,0) \Square{\dotsize}; \draw (2,0) node[below]{$N$};
     \draw (2,0) arc (0:180:1);
    \end{tikzpicture} & 
    \begin{equation}
      \int_{\eta_\text{LSS}}^{\eta_{N+1}}d\eta_N~\nabla_{\hat{\mathbf{n}}}^{i_N}\left[\Psi^\text{NL}_\text{W}(\eta_N)\right]\left(2\frac{r_N - r_\text{LSS}}{r_N r_\text{LSS}}\right)\frac{\partial}{\partial n^{i_N}}\Theta(\hat{\mathbf{n}})\nonumber
    \end{equation}
    or 
    \begin{equation}
      \int^{\eta_0}_0 \text{d}\tilde{\eta}\int_{\tilde{\eta}}^{\eta_{N+1}}d\eta_N~\nabla_{\hat{\mathbf{n}}}^{i_N}\left[\Psi^\text{NL}_\text{W}(\eta_N)\right]\left(2\frac{r_N - \tilde{r}}{r_N \tilde{r}}\right)\frac{\partial}{\partial n^{i_N}}S_T(\tilde{\eta},-\hat{\mathbf{n}}\tilde{r})\nonumber
    \end{equation}
    & ~~The lens at node $N$ distorts the unlensed source, i.e.~lens-source coupling.\\
   \hline ~ & ~ & ~ \\
    \begin{tikzpicture}[baseline=(current bounding box.west)]
     \fill (0,0) \Square{\dotsize}; \draw (0,0) node[below]{$M$};
     \draw (1,0) node[below]{$...$};
     \fill (2,0) \Square{\dotsize}; \draw (2,0) node[below]{$N$};
     \draw (2,0) arc (0:180:1);
    \end{tikzpicture} &
    \begin{equation}
      \int_{\eta_\text{LSS}}^{\eta_{N+1}}d\eta_N~\nabla_{\hat{\mathbf{n}}}^{i_N}\left[\Psi^\text{NL}_\text{W}(\eta_N)\right]\left(2\frac{r_N - r_M}{r_N r_M}\right)\nabla^{i_M}_{\hat{\mathbf{n}}}\left[\frac{\partial}{\partial n^{i_N}}\Psi^\text{NL}_\text{W}(\eta_M)\right]\nonumber
    \end{equation}
    or 
    \begin{equation}
      \int_{\tilde{\eta}}^{\eta_{N+1}}d\eta_N~\nabla_{\hat{\mathbf{n}}}^{i_N}\left[\Psi^\text{NL}_\text{W}(\eta_N)\right]\left(2\frac{r_N - r_M}{r_N r_M}\right)\nabla^{i_M}_{\hat{\mathbf{n}}}\left[\frac{\partial}{\partial n^{i_N}}\Psi^\text{NL}_\text{W}(\eta_M)\right]\nonumber
    \end{equation}
    & ~~The lens at node $N$ distorts the lens at node $M$, i.e.~lens-lens coupling.\\
   \hline
  \end{tabular}
 \end{center}
 \caption{The diagrams with their corresponding represented equations and their physical meanings. In the ``Term'' column, the first choice assumes Approximation \ref{SingleSourceApproH} while the second one does not.} \label{DiagramRuleLen} 
\end{table}

Hence, the prescription to writing down the integrals in \eqref{SingleSourceIwithoutApp6} is to construct \emph{all possible actions from the left to the right}. Physically, an $N$th node acting on an $M$th node for $N>M$ means that the object closer to us at $N$ is lensing the object further away at $M$. In our formalism, $\begin{tikzpicture} \fill (0,0) \Square{\dotsize}; \end{tikzpicture}$ is a lens and thus an overarc between two lenses denotes a \emph{lens-lens coupling}. We emphasize that, although we have considered only the lensing effect here, it is easy to generalize our diagrammatic approach to other types of interaction terms such as the redshift and time-delay effects as discussed in Section \ref{SecDiscussion} by adding more node types.

As an example, we list out all the possible diagrams for the third-, forth- and fifth-order terms of the lensed temperature anisotropies $\tilde{\Theta}$ in Table \ref{AllDiagramsTemp}. In fact, except the upper left diagram in each of Table \ref{3rdIteration}, \ref{4thIteration} and \ref{5thIteration}, all the other diagrams involve the lens-lens couplings. Ignoring these couplings is equivalent to applying Approximation \ref{ApproNolenslens} in Section \ref{HighOrderLensing}.

\begin{table}[htbp]
 \begin{center}
 \begin{center}
  \begin{tabular}{| p{2.4cm} | p{2.4cm} |}
    \hline ~ & ~ \\
     \begin{tikzpicture} \CouplingUpR{2}{0}{0}{1,2,0,0}; \end{tikzpicture} & 
     \begin{tikzpicture} \CouplingUpR{2}{0}{0}{1,1,0,0}; \end{tikzpicture} \\
    \hline
  \end{tabular}
 \end{center}
 \subcaption{The third iteration}\label{3rdIteration}
 \begin{center}
  \begin{tabular}{| p{2.4cm} | p{2.4cm} | p{2.4cm} |}
    \hline ~ & ~ & ~ \\
     \begin{tikzpicture} \CouplingUpR{3}{0}{0}{1,2,3,0}; \end{tikzpicture} & 
     \begin{tikzpicture} \CouplingUpR{3}{0}{0}{1,2,2,0}; \end{tikzpicture} &
     \begin{tikzpicture} \CouplingUpR{3}{0}{0}{1,2,1,0}; \end{tikzpicture} \\
    \hline ~ & ~ & ~ \\
     \begin{tikzpicture} \CouplingUpR{3}{0}{0}{1,1,3,0}; \end{tikzpicture} & 
     \begin{tikzpicture} \CouplingUpR{3}{0}{0}{1,1,2,0}; \end{tikzpicture} &
     \begin{tikzpicture} \CouplingUpR{3}{0}{0}{1,1,1,0}; \end{tikzpicture} \\
    \hline
  \end{tabular}
 \end{center}
 \subcaption{The fourth iteration}\label{4thIteration}
 \begin{center}
  \begin{tabular}{| p{2.4cm} | p{2.4cm} | p{2.4cm} | p{2.4cm} |}
    \hline ~ & ~ & ~ & ~ \\
     \begin{tikzpicture} \CouplingUpR{4}{0}{0}{1,2,3,4}; \end{tikzpicture} & 
     \begin{tikzpicture} \CouplingUpR{4}{0}{0}{1,2,3,3}; \end{tikzpicture} & 
     \begin{tikzpicture} \CouplingUpR{4}{0}{0}{1,2,3,2}; \end{tikzpicture} &
     \begin{tikzpicture} \CouplingUpR{4}{0}{0}{1,2,3,1}; \end{tikzpicture} \\
    \hline ~ & ~ & ~ & ~ \\
     \begin{tikzpicture} \CouplingUpR{4}{0}{0}{1,2,2,4}; \end{tikzpicture} & 
     \begin{tikzpicture} \CouplingUpR{4}{0}{0}{1,2,2,3}; \end{tikzpicture} & 
     \begin{tikzpicture} \CouplingUpR{4}{0}{0}{1,2,2,2}; \end{tikzpicture} &
     \begin{tikzpicture} \CouplingUpR{4}{0}{0}{1,2,2,1}; \end{tikzpicture} \\
    \hline ~ & ~ & ~ & ~ \\
     \begin{tikzpicture} \CouplingUpR{4}{0}{0}{1,2,1,4}; \end{tikzpicture} & 
     \begin{tikzpicture} \CouplingUpR{4}{0}{0}{1,2,1,3}; \end{tikzpicture} & 
     \begin{tikzpicture} \CouplingUpR{4}{0}{0}{1,2,1,2}; \end{tikzpicture} &
     \begin{tikzpicture} \CouplingUpR{4}{0}{0}{1,2,1,1}; \end{tikzpicture} \\
    \hline ~ & ~ & ~ & ~ \\
     \begin{tikzpicture} \CouplingUpR{4}{0}{0}{1,1,3,4}; \end{tikzpicture} & 
     \begin{tikzpicture} \CouplingUpR{4}{0}{0}{1,1,3,3}; \end{tikzpicture} & 
     \begin{tikzpicture} \CouplingUpR{4}{0}{0}{1,1,3,2}; \end{tikzpicture} &
     \begin{tikzpicture} \CouplingUpR{4}{0}{0}{1,1,3,1}; \end{tikzpicture} \\
    \hline ~ & ~ & ~ & ~ \\
     \begin{tikzpicture} \CouplingUpR{4}{0}{0}{1,1,2,4}; \end{tikzpicture} & 
     \begin{tikzpicture} \CouplingUpR{4}{0}{0}{1,1,2,3}; \end{tikzpicture} & 
     \begin{tikzpicture} \CouplingUpR{4}{0}{0}{1,1,2,2}; \end{tikzpicture} &
     \begin{tikzpicture} \CouplingUpR{4}{0}{0}{1,1,2,1}; \end{tikzpicture} \\
    \hline ~ & ~ & ~ & ~ \\
     \begin{tikzpicture} \CouplingUpR{4}{0}{0}{1,1,1,4}; \end{tikzpicture} & 
     \begin{tikzpicture} \CouplingUpR{4}{0}{0}{1,1,1,3}; \end{tikzpicture} & 
     \begin{tikzpicture} \CouplingUpR{4}{0}{0}{1,1,1,2}; \end{tikzpicture} &
     \begin{tikzpicture} \CouplingUpR{4}{0}{0}{1,1,1,1}; \end{tikzpicture} \\
    \hline
  \end{tabular}
 \end{center}
 \subcaption{The fifth iteration}\label{5thIteration}
 \end{center}
 \caption{All the possible couplings of the weak lensing effects on the CMB temperature anisotropies in the third, forth and fifth orders. As an example, \eqref{Theta4Example} demonstrates how to construct the formulae from the middle diagram in the first row of Table \ref{4thIteration}.} \label{AllDiagramsTemp}
\end{table}

To illustrate the rules summarized in Table \ref{DiagramRuleLen}, we demonstrate how to construct the formulae for the middle diagram in the first row of Table \ref{4thIteration}. From the diagram, there are one circle node and three square nodes, i.e.~one source and three lenses. The lenses at node 1 and 2 distort the source while the lens at node 3 distorts the lens at node 1. Based on the rules in Table \ref{DiagramRuleLen}, the integral associated with this diagram is
\begin{eqnarray}\label{Theta4Example}  
  \tilde{\Theta}^{[4]}(\hat{\mathbf{n}}) &\supset&
     \begin{tikzpicture} \CouplingUpR{3}{0}{0}{1,2,2,0}; \end{tikzpicture}  \nonumber\\
  &=& \int^{\eta_{0}}_{\eta_\text{LSS}}\text{d}\eta_3
  \nabla_{\hat{\mathbf{n}}}^{i_3}\left[\Psi^\text{NL}_\text{W}(\eta_3,-\hat{\mathbf{n}}r_3)\right]\bigg\{\nonumber\\
  &~&~~~~~~~~~~~
  \int^{\eta_{3}}_{\eta_\text{LSS}}\text{d}\eta_2
  \nabla_{\hat{\mathbf{n}}}^{i_2}\left[\Psi^\text{NL}_\text{W}(\eta_2,-\hat{\mathbf{n}}r_2)\right]\bigg\{\nonumber\\
  &~&~~~~~~~~~~~~~~~~~~~~~~
  \int^{\eta_{2}}_{\eta_\text{LSS}}\text{d}\eta_1\left(2\frac{r_3-r_1}{r_3~r_1}\right)
  \nabla_{\hat{\mathbf{n}}}^{i_1}\left[\frac{\partial}{\partial n^{i_3}}\Psi^\text{NL}_\text{W}(\eta_1,-\hat{\mathbf{n}}r_1)\right]\bigg\{\nonumber\\
  &~&~~~~~~~~~~~~~~~~~~~~~~~~~~~~~~~~~
  \left(2\frac{r_2-r_\text{LSS}}{r_2~r_\text{LSS}}\right)\left(2\frac{r_1-r_\text{LSS}}{r_1~r_\text{LSS}}\right)\frac{\partial}{\partial n^{i_2}}\frac{\partial}{\partial n^{i_1}}\Theta(\hat{\mathbf{n}})~~~\bigg\}\bigg\}\bigg\}.~~~~~~~~~~~~
\end{eqnarray}

While these diagrams may seem frivolous, its power arises when we want to compute the two-point \emph{correlation functions}, e.g. $\langle \Theta^{[N]}\Theta^{[M]}\rangle$ to all possible actions in the $N$th and $M$th orders. To see that, we need a new rule to encapsulate \emph{contractions} between all possible interaction terms (either \begin{tikzpicture} \fill (0,0) \Square{\dotsize}; \end{tikzpicture} or \begin{tikzpicture} \fill (0,0) circle[radius=\dotsize]; \end{tikzpicture}).

To simplify the numerical calculation, we work in the limit of flat-sky,  use the Limber approximation and consider only the linear part of the Weyl potential\footnote{For simplicity, we omit the superscript [I] from now on.} (i.e.~$\Psi^\text{NL}_\text{W}\rightarrow\Psi_\text{W}^\text{[I]}$). Operationally, this works as follows. Recall that for any function $f(\hat{\mathbf{n}})$, we can perform the Fourier transformation
\begin{eqnarray}
 f(\hat{\mathbf{n}})=\int\frac{d\boldsymbol{\ell}}{2\pi} f(\boldsymbol{\ell}) e^{i\boldsymbol{\ell}\cdot\hat{\mathbf{n}}}.
\end{eqnarray}
Using this as a basis to represent both the temperature anisotropies $\Theta$ and the Weyl potential $\Psi_\text{W}$, it can be shown that
\begin{eqnarray}
 \langle\Theta(\boldsymbol{\ell})\Theta^*(\boldsymbol{\ell}')\rangle &=& \delta(\boldsymbol{\ell}-\boldsymbol{\ell}') C^\Theta_\ell,\label{CTheta} \\
 \langle\Psi_\text{W}(\eta,\boldsymbol{\ell})\Psi^*_\text{W}(\eta',\boldsymbol{\ell}')\rangle &=& \delta(\boldsymbol{\ell}-\boldsymbol{\ell}')\delta(\eta-\eta')C^\Psi_\ell(\eta),\label{CPsi}\label{PsiPsiCorrelation}
\end{eqnarray}
where the angle brackets denote the ensemble averages. The first line above is just the unlensed CMB power spectrum while the second line is the equal time power spectrum of the Weyl potential -- the Dirac delta function $\delta(\eta-\eta')$ arises from the Limber approximation. The Limber approximation performs very well because the Weyl potential varies slowly with the wavenumber $k$. Moreover, we ignore the correlation between the temperature anisotropies and the Weyl potential, i.e.~
\begin{eqnarray}
 \langle\Psi_\text{W}(\eta,\boldsymbol{\ell})\Theta^*(\boldsymbol{\ell}')\rangle &\approx& 0. 
\end{eqnarray}
It is a very good approximation because the CMB temperature anisotropies correlate weakly with the Weyl potential.

Diagrammatically, we represent the correlation function $\langle\hdots\rangle$ by a dotted line connecting two nodes. For example, \begin{tikzpicture} \draw [black,dotted] (0,0) -- (1,0); \fill (0,0) circle[radius=\dotsize]; \fill (1,0) circle[radius=\dotsize];\end{tikzpicture} denotes $\langle\Psi_\text{W}\Psi^*_\text{W}\rangle$.  Due to the presence of the Dirac delta $\delta(\eta-\eta')$ in \eqref{PsiPsiCorrelation}, many possible configurations are forbidden due to the fact that we cannot place the lens further than the lensed object from the observer. In Table \ref{ForbiddenDiagram}, we list out all the forbidden contractions in the diagrams, the corresponding terms in the formula and their physical meanings. 

\begin{table}[htbp]
 \begin{center}
  \begin{tabular}{| p{2.7cm} | p{7.1cm} | p{7.5cm} |}
   \hline Forbidden Contractions & Term & Physical Meaning \\ \hline 
   ~~~~~~\begin{tikzpicture} [baseline=(current bounding box.west)]
    \draw [black,dotted] (0,0) -- (1,0); 
    \fill (0,0) \Square{\dotsize};
    \fill (1,0) circle[radius=\dotsize];
   \end{tikzpicture} &
   $\langle\Psi_\text{W}(\boldsymbol{\ell},\eta)\Theta^*(\boldsymbol{\ell}')\rangle$ &
   The CMB temperature anisotropies correlate weakly with the Weyl potential.\\ \hline
   \begin{tikzpicture} [baseline=(current bounding box.west)]
    \fill (0,0) circle[radius=\dotsize];
    \fill (2,0) circle[radius=\dotsize];
    \fill (0,-1) circle[radius=\dotsize];
    \fill (2,-1) circle[radius=\dotsize];
    \draw (0,0) node[below]{$M$};
    \draw (2,0) node[below]{$N$};
    \draw (0,-1) node[above]{$M'$};
    \draw (2,-1) node[above]{$N'$};
    \draw (1,0) node[below]{$\hdots$};
    \draw (1,-1) node[below]{$\hdots$};
    \draw [black,dotted] (0,0) -- (2,-1);
    \draw [black,dotted] (2,0) -- (0,-1);
   \end{tikzpicture} &
   $\theta(\eta_N-\eta_M)\theta(\eta_{N'}-\eta_{M'})\delta(\eta_N-\eta_{M'})\delta(\eta_{N'}-\eta_M)$ &
   Lens at $\eta_N$ must be placed after any lens at $\eta_M (< \eta_N)$. \\ \hline
   \begin{tikzpicture} [baseline=(current bounding box.west)]
    \fill (0,0) circle[radius=\dotsize];
    \fill (1,0) circle[radius=\dotsize];
    \fill (2,0) circle[radius=\dotsize];
    \draw (0,0) node[below]{$M$};
    \draw (1,0) node[below]{$X$};
    \draw (2,0) node[below]{$N$};
    \draw (0.5,0) node[below]{$\hdots$};
    \draw (1.5,0) node[below]{$\hdots$};
    \draw [black,dotted] (0,0) arc (-180:0:1);
   \end{tikzpicture} &
   $\theta(\eta_N-\eta_X)\theta(\eta_X-\eta_M)\delta(\eta_N-\eta_M)$ &
   Lens at $\eta_X$ cannot be placed before and after the lenses at $\eta_M=\eta_N$ simultaneously. \\ \hline
   \begin{tikzpicture} [baseline=(current bounding box.west)]
    \fill (0,0) circle[radius=\dotsize];
    \fill (2,0) circle[radius=\dotsize];
    \draw (0,0) node[below]{$N$};
    \draw (2,0) node[below]{$N+1$};
    \draw [black] (2,0) arc (0:180:1);
    \draw [black,dotted] (0,0) -- (2,0);
   \end{tikzpicture} &
   $\delta(\eta_{N+1}-\eta_N)(r_{N+1}-r_N)$ &
   No lensing effect if the lens is placed at the same location as the lensed object. \\ \hline
  \end{tabular}
 \end{center}
 \caption{The forbidden contractions for the lensing effects on the power spectrum. $\theta$ is the Heaviside step function.  The Heaviside step functions come from replacing the integration $\int^{\eta_{N+1}}_{\eta_\text{LSS}}\text{d}\eta_N$ with $\int^{\eta_0}_{\eta_\text{LSS}}\text{d}\eta_N\theta(\eta_{N+1}-\eta_N)$.} \label{ForbiddenDiagram}

\end{table}

Because of the forbidden contractions, the lens-lens couplings do not contribute to the 4th-order power spectrum. In order to assess the lens-lens coupling effects, we study the 6th-order power spectrum. We construct all the non-zero configurations for the 6th-order power spectrum as shown in Table \ref{PSLensCoupling}. There exist cancellations between some pairs of the diagrams which are crossed out by the arrows. To understand this, we perform the Fourier transformation such that
\begin{eqnarray}
 (\nabla_{\hat{\mathbf{n}}})_i\approx\frac{\partial}{\partial n^i} \rightarrow i \ell_i.
\end{eqnarray}
In Fourier space, each pair of these diagrams looks exactly the same except that one contains $(\boldsymbol{\ell}_M\cdot\boldsymbol{\ell}_X)\delta(\boldsymbol{\ell}_M+\boldsymbol{\ell}_N)$ while another one contains $(\boldsymbol{\ell}_N\cdot\boldsymbol{\ell}_X)\delta(\boldsymbol{\ell}_M+\boldsymbol{\ell}_N)$. The difference comes from an overarc acting on either one of two correlated lenses. The pairs eliminate because $\boldsymbol{\ell}_M = - \boldsymbol{\ell}_N$. 

\begin{table}[htbp]
 \begin{center}
  \begin{tabular}{| p{2.4cm} | p{12.5cm} |}
   \hline Correlations & Diagrams \\
   ~ & ~ \\
   \hline ~ & ~ \\
   $\langle\Theta^{[3]}\Theta^{[3]*}\rangle$ &
   \begin{tikzpicture} [baseline=(current bounding box.west)]
    \Correlations{2}{2}{0}{0}{1}{2}{-1}{-2};
    \CouplingUp{2}{0}{0}{1,2};
    \CouplingDn{2}{0}{-1}{1,2}; 
    \draw (1,-2.5) node[above]{(1A)};
    
    \Correlations{2}{2}{2.5}{0}{1}{-1}{2}{-2};
    \CouplingUp{2}{2.5}{0}{1,2};
    \CouplingDn{2}{2.5}{-1}{1,2};
    \draw (3.5,-2.5) node[above]{(1B)};
    
    \Correlations{2}{2}{5}{0}{1}{-1}{2}{-2};
    \CouplingUp{2}{5}{0}{1,2};
    \CouplingDn{2}{5}{-1}{1,1}; 
    \draw (6,-2.5) node[above]{(1C)};
    
    \Correlations{2}{2}{7.5}{0}{1}{-1}{2}{-2};
    \CouplingUp{2}{7.5}{0}{1,1};
    \CouplingDn{2}{7.5}{-1}{1,1}; 
    \draw (8.5,-2.5) node[above]{(1D)};
   \end{tikzpicture} \\
   ~ & ~ \\
   \hline ~ & ~ \\
   $\langle\Theta^{[4]}\Theta^{[2]*}\rangle$ &
   \begin{tikzpicture} 
    \Correlations{3}{1}{0}{0}{1}{2}{3}{-1};
    \CouplingUp{3}{0}{0}{1,2,3};
    \CouplingDn{1}{0}{-1}{1}; 
    \draw (1,-2.5) node[above]{(2A)};
    
    \Correlations{3}{1}{2.5}{0}{1}{-1}{2}{3};
    \CouplingUp{3}{2.5}{0}{1,2,3};
    \CouplingDn{1}{2.5}{-1}{1}; 
    \draw (3.5,-2.5) node[above]{(2B)};
    
    \Correlations{3}{1}{5}{0}{1}{-1}{2}{3};
    \CouplingUp{3}{5}{0}{1,2,2};
    \CouplingDn{1}{5}{-1}{1}; 
    \draw (6,-2.5) node[above]{(2C)};
    
    \Correlations{3}{1}{7.5}{0}{1}{-1}{2}{3};
    \CouplingUp{3}{7.5}{0}{1,1,2};
    \CouplingDn{1}{7.5}{-1}{1}; 
    \draw (8.5,-2.5) node[above]{(2D)};
   \end{tikzpicture} \\
   ~ & 
   \begin{tikzpicture} 
    \Correlations{3}{1}{0}{0}{1}{2}{3}{-1};
    \CouplingUp{3}{0}{0}{1,2,2};
    \CouplingDn{1}{0}{-1}{1}; 
    \draw (1,-2.5) node[above]{(3A)};
    
    \Correlations{3}{1}{2.5}{0}{1}{2}{3}{-1};
    \CouplingUp{3}{2.5}{0}{1,2,1};
    \CouplingDn{1}{2.5}{-1}{1};
    \draw (3.5,-2.5) node[above]{(3B)};
    
    \draw [->,black] (0,-2) -- (4.2,0.5);
    \draw [black] (4.2,0.5) node[right]{0};
    
    \Correlations{3}{1}{5}{0}{1}{-1}{2}{3};
    \CouplingUp{3}{5}{0}{1,1,3};
    \CouplingDn{1}{5}{-1}{1}; 
    \draw (6,-2.5) node[above]{(3C)};
   \end{tikzpicture} \\
   \hline ~ & ~ \\
   $\langle\Theta^{[5]}\Theta^{[1]*}\rangle$ &
   \begin{tikzpicture} 
    \Correlations{4}{0}{0}{0}{1}{2}{3}{4};
    \CouplingUp{4}{0}{0}{1,2,3,4};
    \CouplingDn{0}{0}{-1}{1,0,0,0}; 
    \draw (1,-1.5) node[above]{(4A)};
    
    \Correlations{4}{0}{2.5}{0}{1}{2}{3}{4};
    \CouplingUp{4}{2.5}{0}{1,2,3,3};
    \CouplingDn{0}{2.5}{-1}{1,0,0,0};  
    \draw (3.5,-1.5) node[above]{(4B)};
    
    \Correlations{4}{0}{5}{0}{1}{2}{3}{4};
    \CouplingUp{4}{5}{0}{1,2,3,2};
    \CouplingDn{0}{5}{-1}{1,0,0,0}; 
    \draw (6,-1.5) node[above]{(4C)};
    
    \draw [->,black] (2.5,-1) -- (6.8,0.9);
    \draw [black] (6.7,0.9) node[right]{0};
    
    \Correlations{4}{0}{7.5}{0}{1}{2}{3}{4};
    \CouplingUp{4}{7.5}{0}{1,2,2,4};
    \CouplingDn{0}{7.5}{-1}{1,0,0,0}; 
    \draw (8.5,-1.5) node[above]{(4D)};
    
    \Correlations{4}{0}{10}{0}{1}{2}{3}{4};
    \CouplingUp{4}{10}{0}{1,2,1,4};
    \CouplingDn{0}{10}{-1}{1,0,0,0}; 
    \draw (11,-1.5) node[above]{(4E)};
    
    \draw [->,black] (7.5,-1) -- (11.8,0.9);
    \draw [black] (11.7,0.9) node[right]{0};
   \end{tikzpicture} \\
   ~ & 
   \begin{tikzpicture} 
    \Correlations{4}{0}{0}{0}{1}{2}{3}{4};
    \CouplingUp{4}{0}{0}{1,2,2,3};
    \CouplingDn{0}{0}{-1}{1,0,0,0};  
    \draw (1,-1.5) node[above]{(5A)};
    
    \Correlations{4}{0}{2.5}{0}{1}{2}{3}{4};
    \CouplingUp{4}{2.5}{0}{1,2,2,2};
    \CouplingDn{0}{2.5}{-1}{1,0,0,0}; 
    \draw (3.5,-1.5) node[above]{(5B)};
    
    \draw [->,black] (0,-1) -- (4.2,0.6);
    \draw [black] (4.2,0.6) node[right]{0};
    
    \Correlations{4}{0}{5}{0}{1}{2}{3}{4};
    \CouplingUp{4}{5}{0}{1,2,1,3};
    \CouplingDn{0}{5}{-1}{1,0,0,0}; 
    \draw (6,-1.5) node[above]{(5C)};
    
    \Correlations{4}{0}{7.5}{0}{1}{2}{3}{4};
    \CouplingUp{4}{7.5}{0}{1,2,1,2};
    \CouplingDn{0}{7.5}{-1}{1,0,0,0}; 
    \draw (8.5,-1.5) node[above]{(5D)};
    
    \draw [->,black] (5,-1) -- (9.2,0.6);
    \draw [black] (9.2,0.6) node[right]{0};
   \end{tikzpicture} \\
   \hline
  \end{tabular}
 \end{center}
 \caption{All the non-zero configurations for the CMB power spectrum from the next-to-leading order of the weak lensing effect. The dotted lines denote correlations while the arrows indicate the cancellations between some pairs of the diagrams.}\label{PSLensCoupling}
\end{table}

Among the rest of the diagrams in Table \ref{PSLensCoupling}, diagrams (1A), (1B), (2A), (2B) and (4A) do not involve the lens-lens couplings. They correspond to the expanded terms in the Taylor series of $\tilde{\Theta}(\hat{\mathbf{n}})=\Theta(\hat{\mathbf{n}}+\boldsymbol{\alpha})$ which have to be calculated non-perturbatively\cite{LensingFullLewis}. In contrast, other residual terms, which are in Column C and D, contain the lens-lens couplings. We write down the temperature power spectrum for all the residual diagrams in the following
\begin{eqnarray}
 \tilde{C}_\ell^{\Theta\text{(6)}}=\int\frac{d\boldsymbol{\ell}_1 d\boldsymbol{\ell}_2}{(2\pi)^4}\Bigg\{~\mathcal{K}_0(\ell_1,\ell_2)\bigg\{
 &~&\frac{1}{2}(\boldsymbol{\ell}_1\cdot\boldsymbol{\ell})^2(\boldsymbol{\ell}_2\cdot\boldsymbol{\ell})^2 C^\Theta_\ell\nonumber\\
 &-&[\boldsymbol{\ell}_1\cdot(\boldsymbol{\ell}-\boldsymbol{\ell}_1)]^2[\boldsymbol{\ell}_2\cdot(\boldsymbol{\ell}-\boldsymbol{\ell}_1)]^2 C^\Theta_{|\boldsymbol{\ell}-\boldsymbol{\ell}_1|}\nonumber\\
 &+&\frac{1}{2}(\boldsymbol{\ell}_1\cdot\boldsymbol{\ell})^2(\boldsymbol{\ell}_2\cdot\boldsymbol{\ell})^2 C^\Theta_\ell\bigg\}\label{eqn:residualterms1}\\
 +\mathcal{K}_0(\ell_1,\ell_2)\bigg\{&~&[\boldsymbol{\ell}_1\cdot(\boldsymbol{\ell}-\boldsymbol{\ell}_1-\boldsymbol{\ell}_2)]^2[\boldsymbol{\ell}_2\cdot(\boldsymbol{\ell}-\boldsymbol{\ell}_1-\boldsymbol{\ell}_2)]^2 C^\Theta_{|\boldsymbol{\ell}-\boldsymbol{\ell}_1-\boldsymbol{\ell}_2|}\nonumber\\
 &-&[\boldsymbol{\ell}_1\cdot(\boldsymbol{\ell}-\boldsymbol{\ell}_1)]^2[\boldsymbol{\ell}_2\cdot(\boldsymbol{\ell}-\boldsymbol{\ell}_1)]^2 C^\Theta_{|\boldsymbol{\ell}-\boldsymbol{\ell}_1|}\bigg\}\\
 +\mathcal{K}_1(\ell_1,\ell_2)\bigg\{&~&2[\boldsymbol{\ell}_1\cdot(\boldsymbol{\ell}-\boldsymbol{\ell}_1-\boldsymbol{\ell}_2)]^2[\boldsymbol{\ell}_2\cdot(\boldsymbol{\ell}-\boldsymbol{\ell}_1-\boldsymbol{\ell}_2)](\boldsymbol{\ell}_2\cdot\boldsymbol{\ell}_1)C^\Theta_{|\boldsymbol{\ell}-\boldsymbol{\ell}_1-\boldsymbol{\ell}_2|}\nonumber\\
 &-&[\boldsymbol{\ell}_1\cdot(\boldsymbol{\ell}-\boldsymbol{\ell}_1)]^2[\boldsymbol{\ell}_2\cdot(\boldsymbol{\ell}-\boldsymbol{\ell}_1)](\boldsymbol{\ell}_2\cdot\boldsymbol{\ell}_1) C^\Theta_{|\boldsymbol{\ell}-\boldsymbol{\ell}_1|}\nonumber\\
 &-&[\boldsymbol{\ell}_1\cdot(\boldsymbol{\ell}-\boldsymbol{\ell}_1)]^2[\boldsymbol{\ell}_2\cdot(\boldsymbol{\ell}-\boldsymbol{\ell}_1)](\boldsymbol{\ell}_2\cdot\boldsymbol{\ell}_1) C^\Theta_{|\boldsymbol{\ell}-\boldsymbol{\ell}_1|}\bigg\}\\
 +\mathcal{K}_2(\ell_1,\ell_2)\bigg\{&~&[\boldsymbol{\ell}_1\cdot(\boldsymbol{\ell}-\boldsymbol{\ell}_1-\boldsymbol{\ell}_2)]^2(\boldsymbol{\ell}_2\cdot\boldsymbol{\ell}_1)^2 C^\Theta_{|\boldsymbol{\ell}-\boldsymbol{\ell}_1-\boldsymbol{\ell}_2|}\nonumber\\
 &-&[\boldsymbol{\ell}_1\cdot(\boldsymbol{\ell}-\boldsymbol{\ell}_1)]^2(\boldsymbol{\ell}_2\cdot\boldsymbol{\ell}_1)^2 C^\Theta_{|\boldsymbol{\ell}-\boldsymbol{\ell}_1|}\label{eqn:residualterms4}\bigg\}~\Bigg\},
\end{eqnarray}
where the functions $\mathcal{K}_i$ are defined as
\begin{eqnarray}
    \mathcal{K}_i(\ell_1,\ell_2)=\int^{\eta_0}_{\eta_{\text{LSS}}}d\eta_1~4 C^\Psi_{\ell_1}(\eta_1) \left(\frac{r_{\text{LSS}}-r_1}{r_{\text{LSS}}~r_1}\right)^2 \int^{\eta_0}_{\eta_1} d\eta_2~4 C^\Psi_{\ell_2}(\eta_2)
\begin{cases}
    \left(\frac{r_{\text{LSS}}-r_2}{r_{\text{LSS}}~r_2}\right)^2 ,& \text{if } i=0,\\
    \left(\frac{r_{\text{LSS}}-r_2}{r_{\text{LSS}}~r_2}\right)\left(\frac{r_1-r_2}{r_1~r_2}\right) ,& \text{if } i=1,\\
    \left(\frac{r_1-r_2}{r_1~r_2}\right)^2,& \text{if } i=2.
\end{cases}
\end{eqnarray}
Eqns~(\ref{eqn:residualterms1}-\ref{eqn:residualterms4}) correspond to Column A, B, C and D of the residual diagrams in Table \ref{PSLensCoupling} respectively. 

\begin{figure}[ht]
\centering
\includegraphics[scale=0.6]{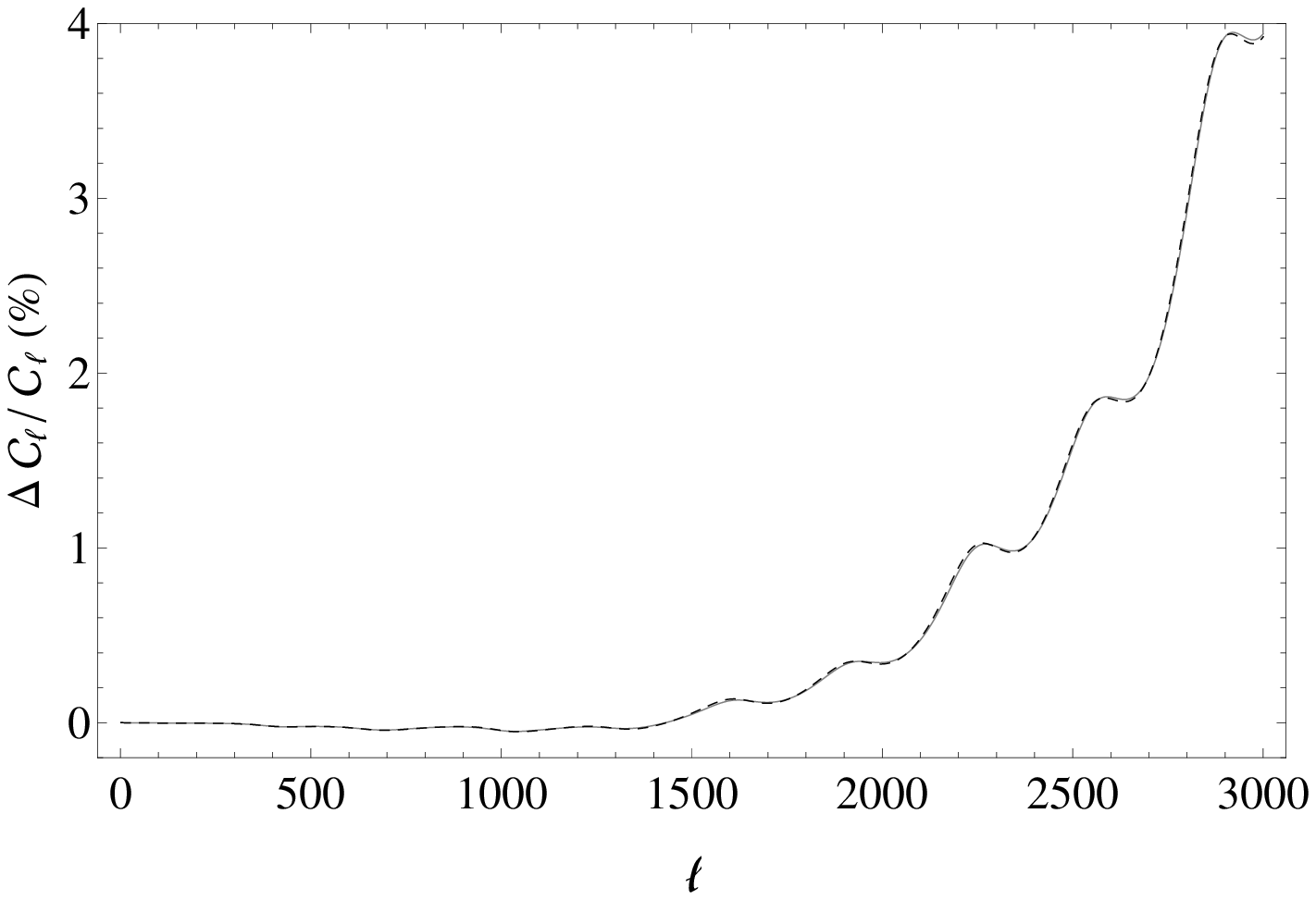}
\includegraphics[scale=0.6]{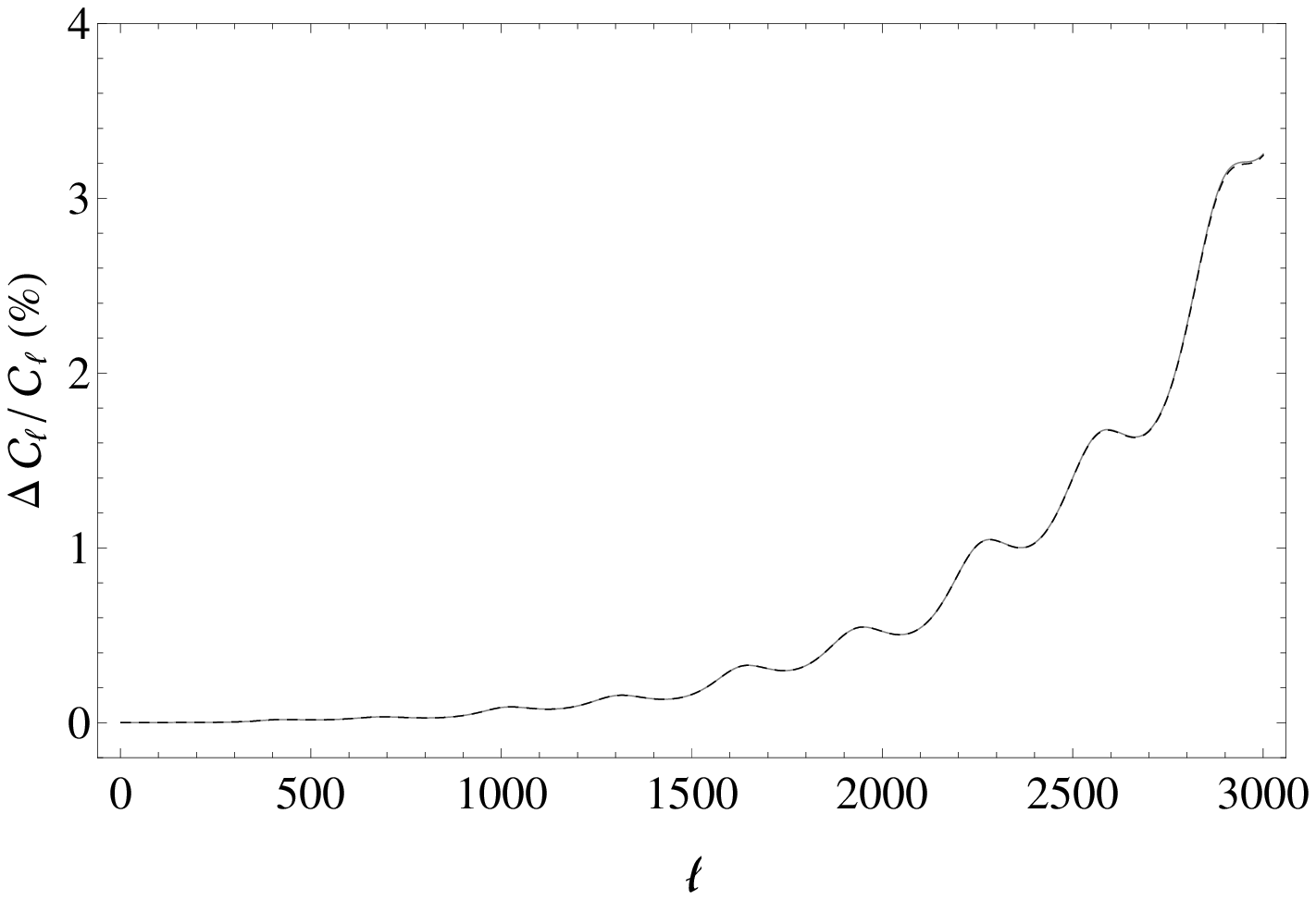}
\caption{The contributions on the temperature power spectrum from the residual diagrams (in Column C and D) containing lens-lens couplings in Table \ref{PSLensCoupling}. The left and right panels show the corrections to the power spectrum from those diagrams in Column C and D of Table \ref{PSLensCoupling} respectively. In the left panel, the solid gray line corresponds to the diagram (1C) while the dashed line corresponds to the diagrams (2C) and (3C)\footnote{Diagrams (2C) and (3C) in Table \ref{PSLensCoupling} have identical contributions and thus we sum them up.}. In the right panel, the solid gray line corresponds to the diagram (1D) while the dashed line corresponds to the diagram (2D). For both panels, the signs of contributions from the diagrams in Row $\langle\Theta^\text{[4]}\Theta^\text{[2]*}\rangle$ are reversed to illustrate the cancellations.}
\label{fig:EachLensLens}
\end{figure}

\begin{figure}[ht]
\centering
\includegraphics[scale=0.8]{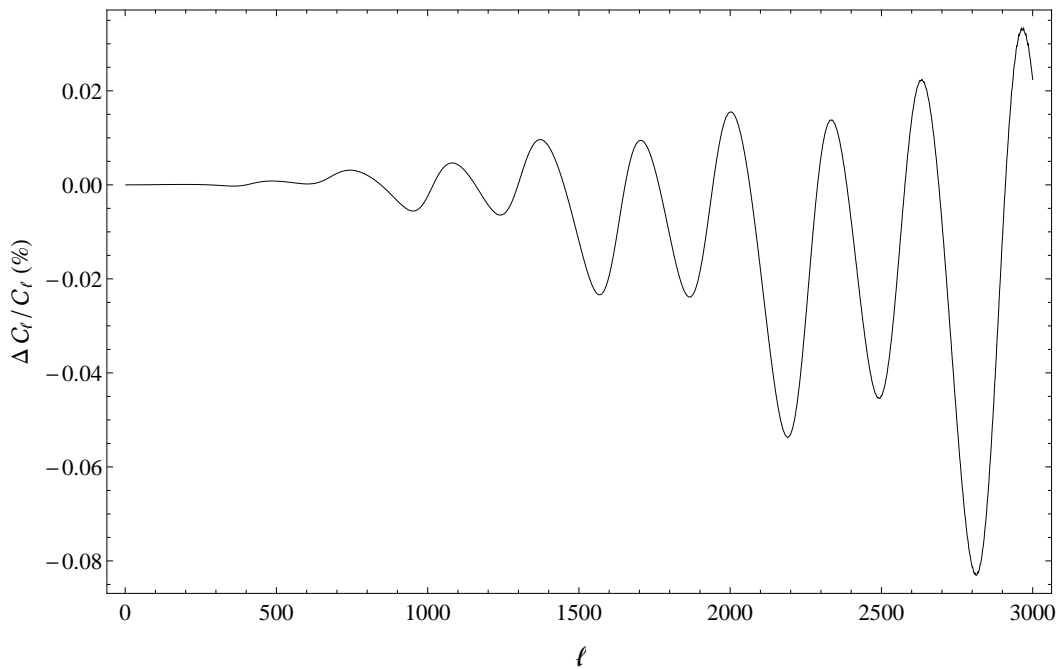}
\caption{The overall correction to the temperature power spectrum from the lens-lens couplings.}
\label{fig:TotalLensLens}
\end{figure}

Since we want to study the effects of lens-lens couplings, we now focus on Column C and D of Table \ref{PSLensCoupling}. In Fig.~\ref{fig:EachLensLens} and \ref{fig:TotalLensLens}, we plot the contributions to the temperature power spectrum from the residual diagrams in Column C and D of Table \ref{PSLensCoupling}. From Fig.~\ref{fig:EachLensLens}, we see that the total contribution of each column in Table \ref{PSLensCoupling} is suppressed by cancellations. The total contribution of each column is at least an order of magnitude smaller than that from an individual diagram in the same column. The cancellations are due to the conservation of the power spectrum from the weak lensing effect. Physically, the positive terms in Eqn.s~(\ref{eqn:residualterms1}-\ref{eqn:residualterms4}) are the power redistributed to the mode $\ell$ while the negative terms are the power redistributed away from the mode $\ell$. 

In summary, as shown in Fig. \ref{fig:TotalLensLens}, the overall correction from the lens-lens couplings is of order $0.1\%$ for $\ell$ up to 3000 and thus is comparable to the cosmic variance. It will be a systematic effect on CMB studies.

\section{Discussion}\label{SecDiscussion}
In Section \ref{ValidApprox}, we calculate the correction from the lens-lens couplings corresponding to Approximation \ref{ApproNolenslens} in Section \ref{SecFormalism}. There are other approximations as mentioned in Section \ref{SecFormalism} but we will leave quantitative assessments of these approximations to the future. Here, we discuss some of the approximations qualitatively. 

Some of these approximations have been studied and their effects on the CMB are believed to be small. The single-source approximation, which is Approximation \ref{SingleSourceAppro2} in Section \ref{Sec2ndLensing} or Approximation \ref{SingleSourceApproH} in Section \ref{HighOrderLensing}, was evaluated in \cite{LensCorrectionFixedSc} using the flat-sky approach. Its corrections to the $TT$ temperature power, the $EE$ polarization power and the $TE$ cross power spectra are of order $0.01\%$ while the correction to the $BB$ polarization power spectrum is about $0.4\%$, for $\ell$ up to $2000$. The pure time-delay effect, included in Approximation \ref{ApproTimeDelay} of Section \ref{HighOrderLensing}, has been studied in \cite{Timedelay} where the effect is estimated to be of order $0.1\%$ correction to the $TE$ cross power spectrum for $\ell \sim 1000$. 

Approximation \ref{ApproPi} in Section \ref{Sec2ndLensing} and Approximation \ref{ApprodSdn} in Section \ref{HighOrderLensing} have not been discussed previously and are first identified in this paper. However, we expect their contributions on the CMB power spectra to be negligible. It is because the corresponding terms do not have the boosting factor $\tilde{r}-r$ (see the argument below \eqref{2ndLensingTemp}) in the coefficients to compensate for the extra order in perturbation. Moreover, the correction to Approximation \ref{ApproPi} in Section \ref{Sec2ndLensing} is proportional to $\ell=2$ multipoles (the term with $\Pi$ in \eqref{2ndLensingTemp}) while the leading correction to Approximation \ref{ApprodSdn} in Section \ref{HighOrderLensing} is in 3rd order as we discussed below \eqref{iterativeI_NCommutedI}. Thus, their corrections should be highly suppressed.

By ignoring $\text{d}x^I/\text{d}\eta$ (Approximation \ref{ApproTimeDelay} in Section \ref{HighOrderLensing}) in the Boltzmann equation, we apply the Born approximation \cite{LensReconPolHirata,LensCorrectionCooray} and neglect the time-delay effect \cite{Timedelay}. Here, we exploit the formalism developed to validate this Approximation to arbitrarily high orders. With the formalism developed in Section \ref{SecFormalism}, we can include these effects by restoring the term
\begin{equation}\label{timedelayterm}
  2 n^i \Psi^\text{NL}_\text{W}\frac{\partial\mathscr{P}_{ab}}{\partial x^I}
\end{equation}
to \eqref{BolzmannEqnLensingOnly}. We can then express the CMB temperature anisotropies as a Dyson series (See Appendix \ref{AppendixA}) with an extra term in the interaction operator of \eqref{InteractionOpLensing}
\begin{eqnarray}\label{InteractionOpTimeDelay}
 \hat{V}_\text{x}(\eta,\hat{\mathbf{n}})&\equiv& n^i\left[\Psi^\text{NL}_\text{W}(\eta,-\hat{\mathbf{n}}r)\right](\hat{\Diamond}_{\hat{\mathbf{n}},r})_i,
\end{eqnarray}
where the subscript x denotes that this term comes from the derivative with respect to $\mathbf{x}$ in the Boltzmann equation and the vector operator $\hat{\Diamond}_{\hat{\mathbf{n}},r}$ is defined such that it is non-zero only when it acts on $\mathcal{X}$
\begin{eqnarray}
 (\hat{\Diamond}_{\hat{\mathbf{n}},r})_i~\mathcal{X}(\eta',-\hat{\mathbf{n}}r')\equiv\frac{2}{r'}\frac{\partial}{\partial n^i} \mathcal{X}(\eta',-\hat{\mathbf{n}}r')
\end{eqnarray}
with $\mathcal{X}$ denoting $\Psi^\text{NL}_\text{W}$ or $S_T$. 

To understand why the lensing effect dominates over the corrections from Born approximation and the time-delay effect, we consider the ratio of the norms of the two interaction operators in \eqref{InteractionOpTimeDelay} and \eqref{InteractionOpLensing}, i.e.~$R\equiv||\hat{V}_\text{x}||/||\hat{V}||$. With the flat-sky approximation, we can replace $\nabla_{\hat{\mathbf{n}}}$ with $\boldsymbol{\ell}$ in Fourier space and have $R\sim r/(r'-r)/\ell$ where $\ell=|\boldsymbol{\ell}|$ and $|n^i|$ is of order $1$. For lens-source couplings which dominate the lensing effect, $r/(r'-r)$ is of order 1 and thus $R \sim 1/\ell$ with $\ell\sim 40$ at the peak of the power spectrum of the lensing potential\footnote{More precisely, lenses at redshifts $z\lesssim 10$ dominate the lensing effect \cite{LensingReportLewis}. For these lenses, $r/(r'-r)\lesssim 2$ and the argument holds as well.}. This estimation is consistent with \cite{Timedelay}. 
Similar argument holds for including redshift effects (Approximation (\ref{ApproRedshift}) in Section \ref{SecFormalism}). This explains why the lensing effect dominates over other effects induced by the perturbed metric. The argument here is made with the interaction operator and is valid throughout the hierarchy, not just in the lowest orders as verified previously in literature. That is, we can safely ignore any high-order couplings involving the redshift and time-delay effects as the pure lensing effect dominates.

Among all the approximations, the Newtonian approximation as shown in Approximation (\ref{ApproNonlinear}) of Section \ref{HighOrderLensing} is the most difficult one to be assessed. On one hand, it relies heavily on the accuracy of the large-scale studies to obtain the non-linear gravitational potentials. On the other hand, the Newtonian approximation performs well in small scales where the evolution is local and the GR effect is negligible. Thus, we can linearize \eqref{NewtonianApprox} and ignore the cross terms. However, the scale of the time integration of the line of sight approach along the lensing effects is comparable to the Hubble radius. The CMB lensing is clearly not a local effect and its GR corrections may be important. Further assessment is needed in the future.

\section{Conclusions and Prospects for Future Work}\label{SecConclusions}
In this paper, we present a new formalism to calculate the lensing effect by directly solving the Boltzmann equation. This allows us to explicitly keep track of \emph{all known physical effects} through the entire time of flight of a CMB photon from recombination to the present observer.  Using this formalism and focusing on temperature anisotropies, we explicitly articulate the approximations required to recover the usual remapping approach used in current studies of the CMB lensing. We discover two new approximations which have not been studied previously although we do not expect that they contribute significantly under the current limits from observations. In addition, we calculate the correction to the CMB temperature power spectrum for the \emph{lens-lens} coupling effects and find that the correction is  $\lesssim 0.1\%$ for $\ell$ up to 3000. It is comparable to the cosmic variance and should be taken into account as systematics. Moreover, since the lensing effect integrates over a time scale 
comparable 
to the Hubble radius, this may raise significant GR corrections which have to be examined in the future. 

Similar formalism can be established for the CMB polarizations. Due to the weak B-mode signals, we expect that the corrections on the B-mode polarization should be stronger compared to that on the CMB temperature in terms of percentage. These corrections may have a significant impact on searches of the primordial B-mode in future experiments, and hence require an accurate assessment.

Finally, we remark that our formalism can be extended to the calculation of weak lensing effects on galaxies by replacing the CMB source terms with the background galaxy power spectrum and hence generalizing the approach of \cite{LensCorrectionHirata}. We will undertake the study of this in future work.

\begin{acknowledgments}
We are grateful for many informative conversations with Anthony Challinor, Ken-ichi Nakao, Lam Hui and Wayne Hu. We would also like to thank Paul Shellard for his generous help and suggestions throughout the project. EAL acknowledges support from an STFC AGP grant ST/L000717/1. The numerical simulations were implemented on the COSMOS supercomputer, part of the DiRAC HPC Facility which is funded by STFC and BIS.
\end{acknowledgments}

\appendix
\section{CMB Lensing Effect as a Dyson Series} \label{AppendixA}
We demonstrate how to formulate the CMB lensing effect into a Dyson series. If we ignore the $\Pi$ term in \eqref{dDeltadn}, we can replace $\tilde{S}_T(\tilde{\eta},\mathbf{k}_1,\hat{\mathbf{n}})$ with $S_T(\tilde{\eta},\mathbf{k}_1)$ and rewrite \eqref{iterativeI_NCommutedI} as 
\begin{eqnarray}
  &~&\frac{1}{(N+1)!}\hat{I}^{[N+1]}(\eta_0,\mathbf{k}_{N+1},\hat{\mathbf{n}})\nonumber\\
  &=&\int^{\eta_0}_0\text{d}\tilde{\eta}~\Bigg\{
  \int^{\eta_{0}}_{\tilde{\eta}}\text{d}\eta_N\left(-\frac{2}{r_N}\right)\int\frac{\text{d}\mathbf{k}'_N\text{d}\mathbf{k}_N}{(2\pi)^\frac{3}{2}}
  \delta(\mathbf{k}_{N+1}-\mathbf{k}'_{N}-\mathbf{k}_{N})
  \nabla_{\hat{\mathbf{n}}}^{i_N}\left[e^{-i\mathbf{k}'_N\cdot\hat{\mathbf{n}}r_N}\Psi^\text{NL}_\text{W}(\eta_N,\mathbf{k}'_N)\right]\bigg\{\nonumber\\
  &~&~~~~~~~~~~~~~~~~~~~~~~~~~~~~~~~~~~~~~~~~~~~~~~~~~~~~~~~~~~~~\vdots\nonumber\\
  &~&~~~~~~~~~~~~~~~~~
  \int^{\eta_{3}}_{\tilde{\eta}}\text{d}\eta_2\left(-\frac{2}{r_2}\right)\int\frac{\text{d}\mathbf{k}'_2\text{d}\mathbf{k}_2}{(2\pi)^\frac{3}{2}}
  \delta(\mathbf{k}_{3}-\mathbf{k}'_{2}-\mathbf{k}_{2})
  \nabla_{\hat{\mathbf{n}}}^{i_2}\left[e^{-i\mathbf{k}'_2\cdot\hat{\mathbf{n}}r_2}\Psi^\text{NL}_\text{W}(\eta_2,\mathbf{k}'_2)\right]\bigg\{\nonumber\\
  &~&~~~~~~~~~~~~~~~~~~~~~~
  \int^{\eta_{2}}_{\tilde{\eta}}\text{d}\eta_1\left(-\frac{2}{r_1}\right)\int\frac{\text{d}\mathbf{k}'_1\text{d}\mathbf{k}_1}{(2\pi)^\frac{3}{2}}
  \delta(\mathbf{k}_{2}-\mathbf{k}'_{1}-\mathbf{k}_{1})
  \nabla_{\hat{\mathbf{n}}}^{i_1}\left[e^{-i\mathbf{k}'_1\cdot\hat{\mathbf{n}}r_1}\Psi^\text{NL}_\text{W}(\eta_1,\mathbf{k}'_1)\right]\bigg\{\nonumber\\
  &~&\bigg[i\sum^{N-1}_{M=1}k'_{M,i_N}(r_N-r_M)+ik_{1,i_N}(r_N-\tilde{r})\bigg]\hdots
  \bigg[ik'_{1,i_2}(r_2-r_1)+ik_{1,i_2}(r_2-\tilde{r})\bigg]
  \bigg[ik_{1,i_1}(r_1-\tilde{r})\bigg]\nonumber\\
  &~&~~~~~~~~~~~~~~~~~~~~~~~~~~~~~~~~~~~~~~~~~~~~~
  e^{-i\mathbf{k}_1\cdot\hat{\mathbf{n}}\tilde{r}}4 S_T(\tilde{\eta},\mathbf{k}_1)~~~~~~~~~~\bigg\}\bigg\}\hdots\bigg\}\Bigg\}.
\end{eqnarray}

With the assumption as shown in \eqref{dSijdn=0}, we can express the photon intensity in configuration space as
\begin{eqnarray}\label{DysonI_N+1}
 &~&\frac{1}{(N+1)!}\hat{I}^{[N+1]}(\eta_0,\hat{\mathbf{n}})=
  \int^{\eta_0}_0\text{d}\tilde{\eta}~\Bigg\{
  \int^{\eta_{0}}_{\tilde{\eta}}\text{d}\eta_N
  \nabla_{\hat{\mathbf{n}}}^{i_N}\left[\Psi^\text{NL}_\text{W}(\eta_N,-\hat{\mathbf{n}}r_N)\right](\hat{\square}_{\hat{\mathbf{n}},r_N})_{i_N}\bigg\{\nonumber\\
  &~&~~~~~~~~~~~~~~~~~~~~~~~~~~~~~~~~~~~~~~~~~~~~~~~~~~~~~~~~~~~~~~~~~~~~~~~~\vdots\nonumber\\
  &~&~~~~~~~~~~~~~~~~~~~~~~~~~~~~~~~~~~~~~~~~~~~~~~~~~~~~
  \int^{\eta_{3}}_{\tilde{\eta}}\text{d}\eta_2
  \nabla_{\hat{\mathbf{n}}}^{i_2}\left[\Psi^\text{NL}_\text{W}(\eta_2,-\hat{\mathbf{n}}r_2)\right](\hat{\square}_{\hat{\mathbf{n}},r_2})_{i_2}\bigg\{ \nonumber\\
  &~&~~~~~~~~~~~~~~~~~~~~~~~~~~~~~~~~~~~~~~~~~~~~~~~~~~~~~~~~~~
  \int^{\eta_{2}}_{\tilde{\eta}}\text{d}\eta_1
  \nabla_{\hat{\mathbf{n}}}^{i_1}\left[\Psi^\text{NL}_\text{W}(\eta_1,-\hat{\mathbf{n}}r_1)\right](\hat{\square}_{\hat{\mathbf{n}},r_1})_{i_1}\bigg\{
  4 S_T(\tilde{\eta},-\hat{\mathbf{n}}\tilde{r})\bigg\}\bigg\}\hdots\bigg\}\Bigg\},~~
\end{eqnarray}
where 
\begin{eqnarray}
 S_T(\tilde{\eta},-\hat{\mathbf{n}}\tilde{r})\equiv\int\frac{\text{d}\mathbf{k}_1}{(2\pi)^\frac{3}{2}}e^{-i\mathbf{k}_1\cdot\hat{\mathbf{n}}\tilde{r}} S_T(\tilde{\eta},\mathbf{k}_1)
\end{eqnarray}
and the vector operator $\hat{\square}_{\hat{\mathbf{n}},r}$ is defined such that it is non-zero only when it acts on $\mathcal{X}$
\begin{eqnarray}\label{SquareOperator}
 (\hat{\square}_{\hat{\mathbf{n}},r})_i~\mathcal{X}(\eta',-\hat{\mathbf{n}}r')\equiv 2 \frac{r-r'}{r~r'}\frac{\partial}{\partial n^i} \mathcal{X}(\eta',-\hat{\mathbf{n}}r')
\end{eqnarray}
with $\mathcal{X}$ denoting $\Psi^\text{NL}_\text{W}$ or $S_T$. Physically, acting $\hat{\square}_{\hat{\mathbf{n}},r}$ on $\Psi^\text{NL}_\text{W}$ corresponds to lens-lens couplings while acting on $S_T$ corresponds to lensing the sources. The diagrammatic approach in Section \ref{DiagramRep} immediately applies to \eqref{DysonI_N+1} by representing $\begin{tikzpicture}[baseline=(current bounding box.west)] \fill (0,0) circle[radius=\dotsize]; \draw (0,0) node[below]{0}; \end{tikzpicture}$ as $S_T$ instead of $\Theta$.

Now, we introduce the \emph{interaction operator} $\hat{V}$ and the evolution operator $\hat{U}$ as
\begin{eqnarray}\label{InteractionOpLensing}
 \hat{V}(\eta,\hat{\mathbf{n}})&\equiv& \nabla_{\hat{\mathbf{n}}}^i\left[\Psi^\text{NL}_\text{W}(\eta,-\hat{\mathbf{n}}r)\right](\hat{\square}_{\hat{\mathbf{n}},r})_i\\
 \hat{U}(\eta_0,\tilde{\eta},\hat{\mathbf{n}})&\equiv& 1 + \sum_{N=1}^\infty\hat{U}_N(\eta_0,\tilde{\eta},\hat{\mathbf{n}}) 
 =\mathcal{T}\left[e^{\int^{\eta_0}_{\tilde{\eta}}\text{d}\eta\hat{V}(\eta,\hat{\mathbf{n}})}\right]
\end{eqnarray}
where $r\equiv\eta_0-\eta$, $\mathcal{T}$ is the time-ordering operator, and
\begin{eqnarray}
 \hat{U}_N(\eta_0,\tilde{\eta},\hat{\mathbf{n}})&\equiv&\frac{1}{N!}\int^{\eta_0}_{\tilde{\eta}}\text{d}\eta_N\hdots
 \int^{\eta_0}_{\tilde{\eta}}\text{d}\eta_{2}\int^{\eta_0}_{\tilde{\eta}}\text{d}\eta_1
 \mathcal{T}\left[\hat{V}(\eta_N,\hat{\mathbf{n}})\hdots \hat{V}(\eta_2,\hat{\mathbf{n}})\hat{V}(\eta_1,\hat{\mathbf{n}})\right].
\end{eqnarray}
Formally, this means that the evolution operator is the solution to the following equation
\begin{equation}
\frac{d}{d\eta}\hat{U}(\eta,\tilde{\eta},\hat{\mathbf{n}}) = \hat{V}(\eta,\hat{\mathbf{n}})\hat{U}(\eta,\tilde{\eta},\hat{\mathbf{n}}).
\end{equation}
Using the evolution operator, \eqref{DysonI_N+1} can be expressed as the following integral
\begin{eqnarray}\label{DysonConfigSpace}
 \frac{1}{(N+1)!}\hat{I}^{[N+1]}(\eta_0,\hat{\mathbf{n}})=4\int^{\eta_0}_0\text{d}\tilde{\eta}~\hat{U}_N(\eta_0,\tilde{\eta},\hat{\mathbf{n}})
 S_T(\tilde{\eta},-\hat{\mathbf{n}}\tilde{r}).
\end{eqnarray}
Thus, the lensed CMB anisotropies observed today are
\begin{eqnarray}\label{lensedCMBDyson}
 \tilde{\Theta}(\hat{\mathbf{n}})=\frac{\hat{I}(\eta_0,\hat{\mathbf{n}})}{4}=\int^{\eta_0}_0\text{d}\tilde{\eta}~\hat{U}(\eta_0,\tilde{\eta},\hat{\mathbf{n}}) S_T(\tilde{\eta},-\hat{\mathbf{n}}\tilde{r}).
\end{eqnarray}
Cast in this form, it is clear that we can interpret \eqref{lensedCMBDyson} as the accumulated lensing effect on the CMB sources at different time $\tilde{\eta}$.

\end{document}